\documentclass[11pt]{article}

\usepackage[a4paper,margin=2.5cm]{geometry}
\usepackage{amsmath,amssymb,bm}
\usepackage{graphicx}
\usepackage{booktabs,array}
\usepackage{caption,subcaption}
\usepackage{xcolor}
\usepackage{microtype}
\usepackage{setspace}
\usepackage{float}
\usepackage[hidelinks]{hyperref}
\usepackage[backend=biber,style=nature,sorting=none,maxbibnames=20]{biblatex}
\newcommand{\PHASE}{\textsc{PHASE}}
\newcommand{\AtwoA}{A$_\mathrm{2A}$}

\newcommand{\figorplaceholder}[3][0.94\linewidth]{%
  \IfFileExists{#2}{\includegraphics[width=#1]{#2}}{%
  \fbox{\parbox[c][0.15\textheight][c]{#1}{\centering\textbf{Figure placeholder}\\[0.5em]#3}}}}

\makeatletter
\def\moverlay{\mathpalette\mov@rlay}
\def\mov@rlay#1#2{\leavevmode\vtop{%
   \baselineskip\z@skip \lineskiplimit-\maxdimen
   \ialign{\hfil$\m@th#1##$\hfil\cr#2\crcr}}}
\newcommand{\charfusion}[3][\mathord]{
    #1{\ifx#1\mathop\vphantom{#2}\fi
        \mathpalette\mov@rlay{#2\cr#3}
      }
    \ifx#1\mathop\expandafter\displaylimits\fi}
\makeatother

\newcommand{\bigcupdot}{\charfusion[\mathop]{\bigcup}{\cdot}}

\title{\textbf{PHASE: encoding global protein ensembles with local Hamiltonians and all-atom backmapping}}

\author{Daniele Angioletti$^{1}$ \and Marco Nobile$^{1}$ \and Matteo Carli$^{1}$ \and Vittorio Limongelli$^{1,*}$}
\date{}

\begin{document}
\maketitle

\begin{center}
\small
$^{1}$Euler Institute, Faculty of Biomedical Sciences, Universit\`a della Svizzera italiana, 6900 Lugano, Switzerland\\
$^*$Correspondence: vittoriolimongelli@gmail.com
\end{center}

\begin{abstract}
Protein function is governed by conformational ensembles, which can be viewed as high-dimensional probability distributions over molecular conformations. Yet the statistical organization of these distributions is often represented only implicitly, either through collections of simulation trajectories or within high-capacity generative models. Here, we introduce \textbf{\PHASE{}} (Protein Hamiltonians for Sampling of Ensembles), \textbf{a system-specific framework that converts atomistic conformational ensembles into an explicit and interpretable statistical model}. Applied to ten conformational ensembles derived from approximately 37~$\mu$s of atomistic simulations of the adenosine \AtwoA{} receptor, Hamiltonians containing only local residue couplings within 6~\AA{} reproduce residue-wise and pairwise microstate statistics, including correlations between residues that are not directly coupled in the model. Moreover, independently fitted inactive and active reference Hamiltonians define an endpoint preference coordinate that organizes newly sampled ligand-, effector- and conformation-dependent ensembles along the \AtwoA{} activation landscape without receiving these biochemical labels as model inputs. Finally, a cluster-conditioned all-atom reconstruction model preserves the prescribed residue microstate patterns of newly sampled configurations, closing the coarse-graining--sampling--backmapping cycle. The resulting discrete representation additionally admits direct QUBO encoding, enabling classical annealing and providing a route toward future quantum-annealing implementations. \PHASE{} therefore provides a protein-general procedure for constructing compact, interpretable and atomistically realizable statistical models of protein conformational ensembles.
\end{abstract}

\section{Introduction}

Protein function is often better described by a conformational ensemble than
by a single structure.
Proteins populate multiple conformational states on complex energy landscapes, and ligand binding, catalysis, allostery and molecular recognition can depend on shifts in the populations of these states rather than on a unique structural rearrangement \cite{Frauenfelder1991Landscape,HenzlerWildman2007Dynamic,
Boehr2009Ensembles,Motlagh2014Allostery}.

This ensemble view is particularly evident for G protein-coupled receptors (GPCRs), where ligands and intracellular effectors reshape a multidimensional conformational landscape associated with different functional responses.
For the adenosine \AtwoA{} receptor, spectroscopic experiments directly resolve coexisting inactive and active conformations whose populations are shifted by inverse agonists, agonists and G-protein coupling, while atomistic simulations reveal the distributed residue rearrangements connecting these states \cite{DAmore2024Chem,Ye2018A2A,Fernandes2021A2A}.
Describing such systems therefore requires not only structurally diverse conformations, but also a statistical model of how those conformations are populated and coupled.

Molecular dynamics (MD) provides a natural route to such information by sampling protein conformations under an explicit molecular Hamiltonian.
However, the raw output of a MD trajectory is simply a finite collection of temporally correlated frames, rather than a compact probabilistic model of the sampled ensemble.
Statistical models such as Markov state models address this problem by discretizing conformation space and learning transitions between states, enabling thermodynamic and kinetic information to be reconstructed from simulation data \cite{Noe2009PNAS,Prinz2011JCP}.
Their states, however, typically describe global molecular conformations.
As system size and conformational complexity increase, representing the molecule through a combinatorial set of coupled local variables provides an attractive alternative to enumerating global states.
This perspective has been formalized in the framework of Markov field models, which decompose large molecular systems into interacting local sites and explicitly identify Ising, Potts and graphical models as possible statistical representations \cite{Hempel2022MarkovFields}.

In parallel, deep generative learning has rapidly expanded the possibilities for molecular ensemble generation.
Boltzmann Generators demonstrated that system-specific neural transformations can directly sample equilibrium distributions \cite{Noe2019Boltzmann}, while more recent approaches such as Distributional Graphormer and BioEmu amortize ensemble generation across systems through conditioning on molecular
descriptors such as protein sequence \cite{Zheng2024DiG,Lewis2025BioEmu}.
Other approaches increasingly operate at high structural resolution: PepFlow directly samples all-atom peptide conformations \cite{Abdin2024PepFlow}, aSAM learns heavy-atom protein ensembles from MD in a latent representation \cite{Janson2025aSAM}, and LD-FPG demonstrates system-specific full-atom conformational generation from MD trajectories of a GPCR \cite{Sengar2025LDFPG}.
These developments also make clear that structural diversity alone is not a
complete description of a protein ensemble: the generated conformations must
carry meaningful statistical weights, and the organization of those
probabilities must itself be accessible to analysis.
A complementary challenge is therefore to represent an ensemble in a compact form that exposes its statistical dependencies and can be directly interrogated and manipulated.
In high-capacity generative models, this information is generally distributed
across many neural parameters; for mechanistic analysis, it can instead be
useful to \textbf{expose an explicit energy model} whose contributions can be associated with individual residues and residue pairs.

\textbf{Pairwise Potts models provide such a representation}.
In protein science they have been used predominantly in sequence space, where each site represents an amino-acid position and its categorical states correspond to residue identities.
Maximum-entropy and direct-coupling approaches showed that single-site fields and pairwise couplings inferred from multiple-sequence alignments recover residue conservation, coevolution and structural contacts \cite{Marks2011DCA,Ekeberg2013PLM}.
Potts Hamiltonians \cite{Potts_1952,PottsReview1982} have subsequently been used to study sequence fitness and protein design, and recent neural approaches predict structure-conditioned Potts energy functions directly from protein backbones \cite{Li2023Potts,Birnbaum2026PottsMPNN}.
PottsMPNN, in particular, demonstrates that retaining an explicit decomposition into single-residue and pairwise energy terms can improve modeling of the sequence--energy landscape even in modern graph-neural-network protein design \cite{Birnbaum2026PottsMPNN}.
These developments motivate repurposing the same statistical form \emph{from sequence variation to conformational variation}.

Here we introduce \textbf{\PHASE} (Protein Hamiltonians for Sampling of Ensembles), a system-specific framework that uses a \textbf{Potts Hamiltonian} to model the \emph{conformational microstates} occupied by individual residues across an ensemble of a given protein system.
Each protein conformation is projected onto a shared discrete representation in which every residue occupies one of $K_r$ microstates obtained by clustering its local backbone and side-chain torsional conformation.
The resulting protein conformation $\mathbf{x}=(x_1,\ldots,x_N)$ is assigned an explicit statistical energy by a pairwise Potts Hamiltonian.
Importantly, pairwise couplings are restricted to a sparse graph of residues that become spatially proximal in the atomistic ensemble.
For the protein ensembles examined here, \PHASE{} demonstrates that a sparse Hamiltonian with interactions confined to local structural neighbourhoods is sufficient to recover the global statistical organization of the reduced conformational ensemble.

The resulting Hamiltonian is not intended as a replacement for the underlying atomistic force field.
Its fields and couplings are effective statistical parameters obtained by fitting the observed frequencies of the states at the chosen residue-level representation.
This distinction is also what makes the model useful: the Hamiltonian can be evaluated directly, sampled through Gibbs and replica-exchange Monte Carlo, compared across different molecular conditions, and mapped to a binary quadratic representation for annealing-based optimization.
Since reducing a protein to residue microstates removes the atomic information required for structural applications, \PHASE{} couples this statistical model to a cluster-conditioned structure generator that reconstructs sampled discrete configurations at all-atom resolution.
The statistical representation and the geometric reconstruction are therefore separated: the former models the probabilities of the combinations of coarse-grained microstates, whereas the latter generates atomic-resolution molecular structures from those combinations.

\textbf{\PHASE{}} thus builds a discrete, sparse, interpretable generative model from an atomistic trajectory ensemble, in which locality in the three-dimensional space is encoded through the coupling graph of a Potts Hamiltonian.
It \textbf{is a coarse-grained generative model} that can be \textbf{evaluated, sampled} and \textbf{interrogated} directly, and \textbf{backmapped generatively} to continuous atomistic molecular structures.

We apply \PHASE{} to ten \AtwoA{} conformational ensembles derived from eight atomistic trajectories spanning inactive and active receptor states, inverse-agonist, neutral-antagonist and agonist conditions, and the presence or absence of mini-G$_s$.\cite{DAmore2024Chem}

\section{Results}

\subsection{\PHASE{}: representing ensembles in a shared residue-state space}

\begin{figure}[ht]
\captionsetup{font={small,stretch=1.0}}
\centering
\includegraphics[width=\textwidth]{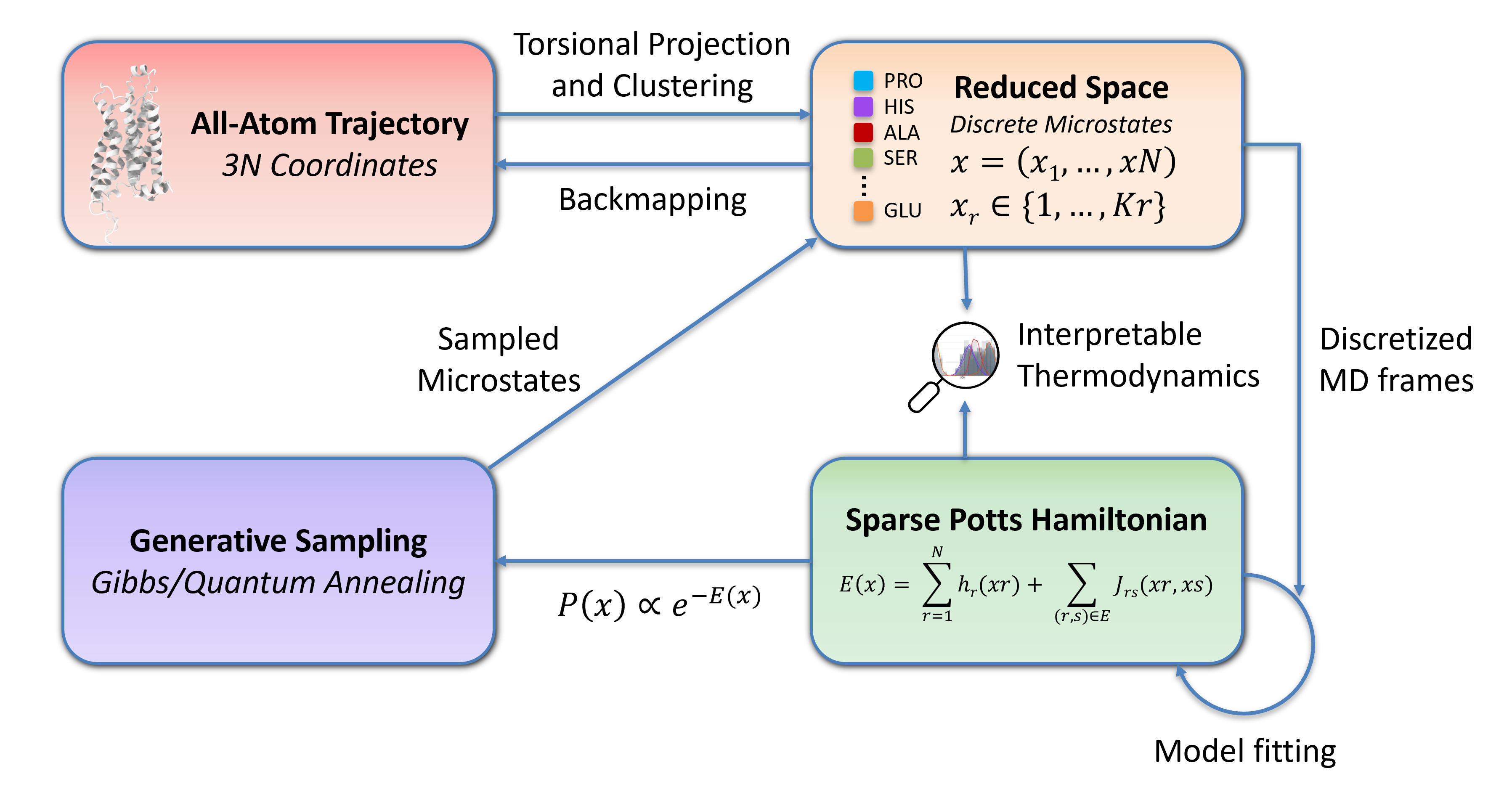}
\caption{\textbf{PHASE constructs a closed statistical representation of
protein conformational ensembles.}
Atomistic trajectories of a protein under one or more conditions are projected onto a shared residue-level torsional microstate space.
Sparse Potts Hamiltonians encode and sample the statistical organization of conformations in this reduced representation.
A cluster-conditioned backmapping model then reconstructs sampled microstate configurations at all-atom resolution, completing the representation--sampling--reconstruction cycle.}
\label{fig:phase_overview}
\end{figure}

\PHASE{} provides a system-specific framework for converting atomistic protein ensembles into an explicit statistical model that can be sampled and reconstructed at all-atom resolution (Fig.~\ref{fig:phase_overview}).
The method can be applied to any protein system for which suitable atomistic ensembles are available.

The first stage of \PHASE{} constructs a shared discrete representation of the
protein conformational ensemble.
For each residue, torsional conformations observed across the selected reference ensembles are pooled and clustered into a small set of recurrent microstates.
Clustering is performed independently for every residue, so that each discrete label retains the same structural meaning across all conditions included for that protein.
An atomistic conformation is consequently represented as $\mathbf{x}=(x_1,\ldots,x_N)$, where $x_r$ identifies the torsional microstate occupied by residue $r$.
Different ensembles therefore share the same discrete configuration space and differ in the populations and correlations of these residue states.

The second stage assigns statistical weights to this reduced space.
Residue interactions are defined through a sparse contact graph constructed from the same conformational data, and ensemble-specific Potts Hamiltonians are fitted on this common representation.
The resulting Hamiltonians can be interrogated directly and sampled to generate new combinations of residue microstates according to the learned distribution.

The final stage maps these sampled discrete configurations back to atomistic resolution.
A cluster-conditioned backmapping model reconstructs all-atom coordinates while conditioning each residue on the microstate specified by the sampled \PHASE{} configuration.
Sampling therefore produces not only discrete state vectors but candidate atomistic conformations whose consistency with the prescribed microstates and structural quality can be evaluated directly.
This establishes a continuous workflow from atomistic ensembles, through an explicit and manipulable statistical representation, to newly generated all-atom structures.

We apply this framework to the study of \AtwoA{} GPCR, using atomistic MD trajectories spanning distinct receptor starting conformations, ligand conditions and mini-Gs coupling (Table~\ref{tab:a2a_systems}).
These trajectories define ten analysis ensembles because the 5~$\mu$s active-starting apo trajectory is split
into two temporally distinct conformational regimes.

The resulting dataset contains approximately 37.26~$\mu$s of atomistic MD.
INzma and Theo contribute 1.26 and 1.0~$\mu$s, respectively, whereas INapo, INeca, the complete FApo/pAs trajectory, FApoG, FAN and FANG contribute 5~$\mu$s each \cite{DAmore2024Chem}.
All ensembles contribute to the common \AtwoA{} residue-state basis and contact graph, providing a single representation in which conformations sampled under different biochemical conditions can be compared directly.

Importantly, because the statistical parameters reflect the conformational distributions generated by classical MD, the resulting models inherently reflect the finite-time sampling and metastable confinement of the underlying trajectories.
While this restricts the support of each Hamiltonian to the conformational basins explored during simulation, this kinetic separation is precisely what enables independently fitted models to capture distinct, condition-dependent metastable states and serve as reference endpoints across the activation landscape.

\begin{table*}[t]
\captionsetup{font={small,stretch=1.0}}
\centering
\small
\begin{tabular}{lllllp{0.25\textwidth}}
\toprule
\textbf{Ensemble} &
\textbf{Conformation} &
\textbf{Ligand} &
\textbf{mini-Gs} &
\textbf{Duration} &
\textbf{Role in PHASE} \\
\midrule

INzma &
inactive &
ZMA &
no &
1.26~$\mu$s &
Inverse-agonist-stabilized inactive endpoint. \\

INapo &
inactive &
none &
no &
5.0~$\mu$s &
Inactive-starting apo ensemble; used for locality and sampling validation. \\

Theo &
inactive &
theophylline &
no &
1.0~$\mu$s &
Neutral-antagonist-bound ensemble. \\

INeca &
inactive &
NECA &
no &
5.0~$\mu$s &
Agonist-bound, inactive-starting ensemble. \\

pAs &
active &
none &
no &
3.5~$\mu$s &
Post-transition segment of the same trajectory, representing the stabilized
pseudo-active ensemble. \\

ACzma &
active &
ZMA &
no &
5.0~$\mu$s &
Active-starting ensemble containing the inverse agonist ZM241385; used to distinguish conformational organization from ligand identity. \\

FApo &
active &
none &
no &
1.5~$\mu$s &
Initial segment of the active-starting apo trajectory, before stabilization
of the pseudo-active state. \\

FApoG &
active &
none &
yes &
5.0~$\mu$s &
Apo active-starting ensemble coupled to mini-Gs. \\

FAN &
active &
NECA &
no &
5.0~$\mu$s &
Agonist-bound active-starting ensemble without mini-Gs. \\

FANG &
active &
NECA &
yes &
5.0~$\mu$s &
Agonist- and mini-Gs-stabilized active endpoint. \\

\bottomrule
\end{tabular}

\caption{\textbf{\AtwoA{} conformational ensembles represented with PHASE.}
Nine underlying MD trajectories define the ten analysis ensembles listed here.
All ensembles are projected onto the same residue-wise microstate basis and contact graph.
\PHASE{} Hamiltonians contain only receptor microstate variables; ligand identity, pharmacological annotation, starting-state identity and mini-Gs coupling are not supplied as
model inputs.}
\label{tab:a2a_systems}
\end{table*}

\subsection{Local residue couplings reproduce global reduced-space organization}

The \PHASE{} Hamiltonian is intentionally restricted to interactions between residues that become spatially proximal in the MD ensembles.
It is therefore relevant to quantify how local this graph can remain while still reproducing the statistics of the complete reduced protein ensemble.
We examine this trade-off using INapo as a representative trajectory and fit otherwise equivalent Potts models with $C_\alpha$ contact cutoffs of 4, 5, 6, and 10~\AA{}.

\begin{figure}[ht]
    \captionsetup{font={small,stretch=1.0}}
    \vspace{-20pt}
    \centering
    \includegraphics[width=0.9\linewidth]{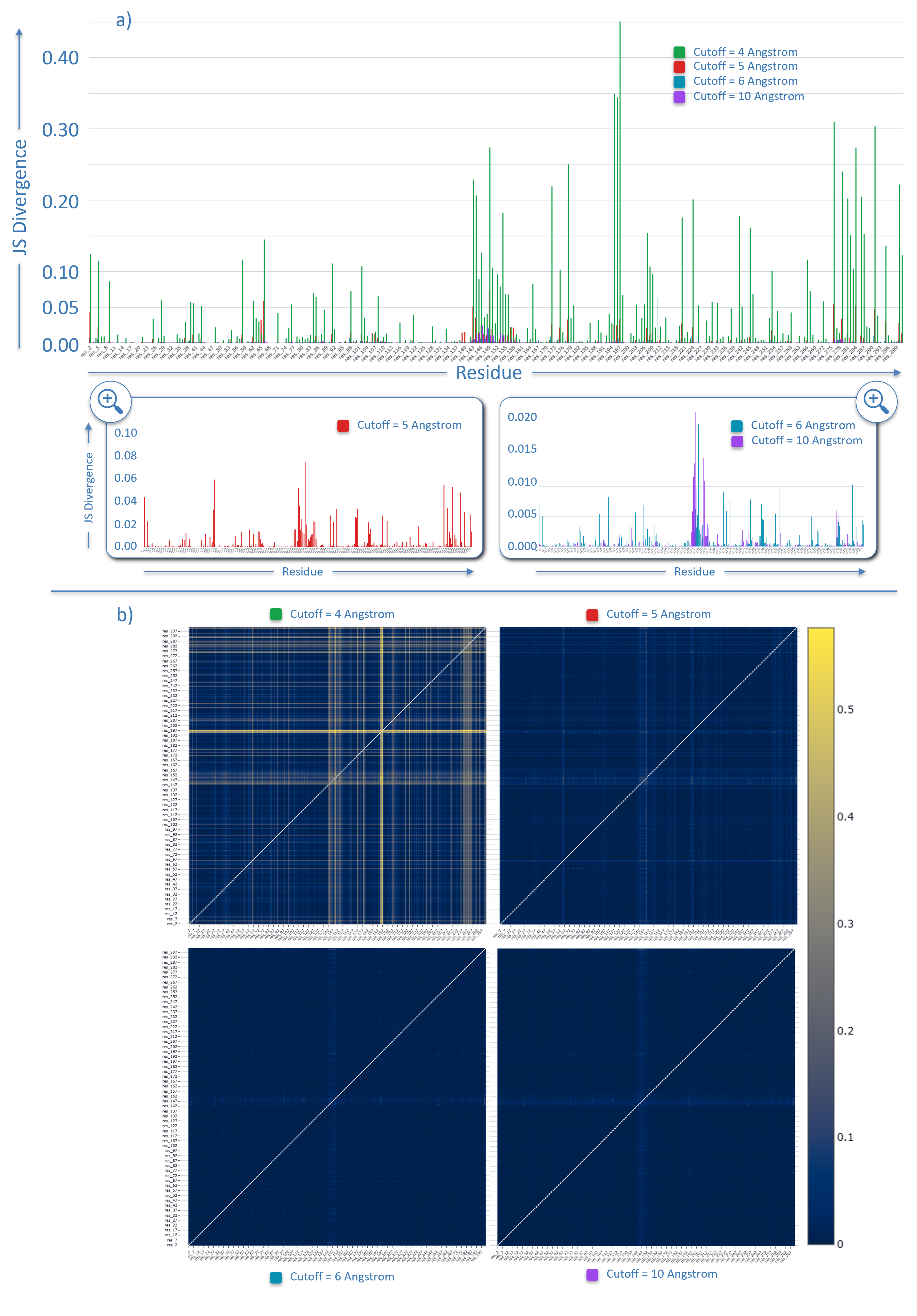}
    \caption{\textbf{Local Potts couplings reproduce global pairwise statistics of the INapo reduced ensemble.}
    \textbf{(a)} Residue-wise Jensen--Shannon (JS) divergence between MD and Potts microstate distributions across $C_\alpha$ contact cutoffs (4, 5, 6, 10\,\AA); bottom insets magnify lower divergence ranges. 
    \textbf{(b)} All-to-all residue-pair JS divergence matrices across the same cutoffs, evaluating all pairs regardless of direct graph connectivity. Lower JS indicates closer agreement.}
    \label{fig:locality_validation}
\end{figure}

First, in Fig.~\ref{fig:locality_validation}a we compare the microstate population of every residue between the MD trajectory and samples generated from each fitted Hamiltonian via Gibbs sampling ~\cite{CasellaGeorge1992} (see \ref{subsec:replica_exchange_sampling} for details).
As reproducing individual residue microstate populations does not establish that the model captures how residue states are organized jointly across the receptor, we perform a more stringent test by comparing the joint microstate distribution of \emph{every} residue pair between MD and Potts samples (Fig.~\ref{fig:locality_validation}b).
Crucially, these matrices include all residue pairs, not only pairs connected by an explicit edge in the fitted Potts graph.

The two analyses show the same clear dependence on graph locality.
At a 4~\AA{} cutoff, the model fails to reproduce both residue-wise and pairwise statistics, with several residues showing large Jensen-Shannon (JS) divergences in their marginal populations and widespread, strongly structured discrepancies across the all-pair JS matrix.
Increasing the cutoff to 5~\AA{} substantially reduces both types of error, although residue-specific and pairwise deviations remain.
At 6~\AA{}, residue marginals are reproduced with consistently low JS divergence and the all-pair JS matrix becomes uniformly low.
Increasing the cutoff further to 10~\AA{} yields only limited additional improvement in either measure, with pairwise statistics closely resembling those obtained at 6~\AA{}. Thus, statistical fidelity largely saturates at a contact cutoff of approximately 6~\AA{}.

This result is important because residue pairs separated beyond the 6~\AA{} cutoff, thus not directly connected in the contact graph, have no direct coupling term $J_{rs}$ in the Hamiltonian.
Nevertheless, their joint state distributions are recovered by samples from the model.
The interaction graph can therefore remain spatially local while samples from the resulting joint distribution reproduce nonlocal pairwise correlations observed in MD.
These correlations arise through the connected network of local statistical dependencies, without requiring direct coupling terms.

We therefore adopt 6~\AA{} as the contact cutoff for the subsequent analyses,
as it provides a simpler and faster-to-train model, while retaining essentially the statistical fidelity obtained with the denser 10~\AA{} model.
Additional residue-wise validation at the selected 6~\AA{} cutoff is shown for the INzma, FANG and INeca ensembles in Supplementary Fig.~\ref{fig:supp_key_models_js_r6}, supporting the use of the same graph construction across inactive, active and agonist-bound conditions.

To further verify that this agreement does not arise from reproducing the specific configurations used for fitting, we perform a temporally disjoint validation on INapo.
A Potts Hamiltonian fitted only to the first trajectory block generates configurations whose distance from the training ensemble is comparable to that of the unseen second block. This suggests that the model has learned a coarse-grained representation  of the inactive ensemble, rather than simply memorizing the discrete configurations observed during fitting.
Pseudo-active and active conformations remain instead progressively more distant (Supplementary Fig.~\ref{fig:supp_temporal_validation}).

Together, these analyses establish that sampling from the local \PHASE{} Hamiltonians recovers both residue-level and whole-configuration statistics of the source ensembles. We next show that this statistical fidelity also preserves the higher-level energetic relationships between independently modelled receptor states.

\subsection{PHASE samples preserve ligand-, effector- and conformation-dependent endpoint ordering}
\label{sec:endpoint_ordering}

The shared residue-state representation allows configurations generated from independently fitted \PHASE{} Hamiltonians to be evaluated using the same statistical coordinates. We exploit this property to test whether sampling from the state-specific models preserves the organization of the original \AtwoA{} ensembles along the activation landscape.

For each investigated condition, we generate configurations by replica-exchange Gibbs sampling from the corresponding 6~\AA{} Hamiltonian.
Each sampled configuration is then evaluated against two well-stabilized reference endpoints.
The inactive reference Hamiltonian is fitted to INzma, which combines an inactive receptor with the inverse agonist ZM241385, whereas the active reference is fitted to FANG, which combines the active receptor, the full agonist NECA and mini-Gs.
These conditions provide deliberately stringent references for the two ends of the activation landscape: ZM241385 preferentially stabilizes inactive \AtwoA{} conformations, while agonist binding and G-protein coupling favour increasingly active receptor states
\cite{Ye2016Nature,Carpenter2016Nature,Lee2019Structure,DAmore2024Chem}.

After zero-sum gauge normalization, each reduced configuration $\mathbf{x}$ is evaluated through the \textit{endpoint preference score}

\begin{equation}
\Delta E_{\mathrm{INzma-FANG}}(\mathbf{x})
=
\widetilde E_{\mathrm{INzma}}(\mathbf{x})
-
\widetilde E_{\mathrm{FANG}}(\mathbf{x}).
\label{eq:delta_endpoint_results}
\end{equation}

Increasing values correspond to progressively greater relative preference for the FANG Hamiltonian.
Across the complete receptor, the resulting distributions span a broad inactive--active range (Fig.~\ref{fig:regional_endpoint}a, top), with their central positions approximately ordered as

\begin{equation}
\mathrm{INzma}
<
\mathrm{INapo}
<
\mathrm{Theo}
<
\mathrm{INeca}
<
\mathrm{pAs}
<
\mathrm{ACzma}
<
\mathrm{FApoG}
<
\mathrm{FAN}
<
\mathrm{FApo}
<
\mathrm{FANG}.
\label{eq:endpoint_ordering}
\end{equation}

Neighbouring distributions overlap substantially, and this sequence should therefore be understood as an ensemble ordering rather than as a discrete classification.

Importantly, these distributions are obtained from newly sampled configurations rather than from the MD frames used to fit the models.
Direct comparison with endpoint-score distributions evaluated on the corresponding MD ensembles shows strong agreement across conditions (Supplementary Fig.~\ref{fig:supp_md_vs_sampling_energies}), demonstrating that sampling preserves this higher-level organization in addition to the marginal statistics established above.

The broad progression is consistent with established \AtwoA{} conformational pharmacology.
Removal of ZM241385 shifts INapo away from the fully inactive reference, consistent with the known constitutive basal activity of unliganded \AtwoA{} and its spontaneous exploration of intermediate substates in the absence of an inverse agonist \cite{Ye2016Nature,DAmore2024Chem}.
Concurrently, the neutral antagonist theophylline remains on the inactive side but is less strongly biased toward INzma than ZM241385.
Introducing the agonist NECA into an inactive-starting receptor shifts INeca further toward the active reference, consistent with agonist-induced enrichment of active conformations, while its separation from FANG agrees with the observation that agonist-bound receptor without a transducer can retain intermediate intracellular geometries \cite{Ye2016Nature,Lebon2011Nature,Lee2019Structure}.
The independently characterized pAs ensemble occupies an intermediate region of the same coordinate, consistent with its identification as a metastable pseudo-active state along the \AtwoA{} activation landscape \cite{DAmore2024Chem}.
On the active side, FApoG and FAN show strong FANG preference under mini-Gs coupling or agonist binding, respectively, whereas their combination in FANG defines the fully stabilized active reference \cite{Carpenter2016Nature,Lee2019Structure,DAmore2024Chem}.

Two comparisons show more directly that the endpoint score is not simply recovering ligand or effector labels.
First, the sampled ACzma ensemble remains on the active side of the intermediate ensembles despite originating from a trajectory containing the same inverse agonist as INzma but starting from an active receptor conformation.
This ligand--conformation mismatch reflects the receptor states actually visited during the trajectory, rather than the pharmacological identity of ZM241385 alone \cite{DAmore2024Chem}.

Second, FApo and pAs provide a particularly stringent within-trajectory control.
Their source ensembles correspond to successive portions of the same apo trajectory and therefore share identical ligand and effector conditions.
Nevertheless, independently sampling the Hamiltonians fitted to these two segments preserves their marked separation along the endpoint coordinate: FApo retains strong FANG preference, whereas pAs shifts toward the centre of the coordinate.
\PHASE{} therefore reproduces two distinct conformational regimes learned from different temporal portions of the same receptor
trajectory.

Together, these results show that the endpoint Hamiltonians capture the distributed conformational response of the receptor to ligand binding, effector coupling, starting-state mismatch and spontaneous relaxation.

We note that $\Delta E_{\mathrm{INzma-FANG}}$ is a relative statistical preference score rather than a physical free-energy difference between inactive and active macrostates.
Since the partition functions of the two independently fitted Hamiltonians are not evaluated, the absolute zero of the score cannot be interpreted as equal endpoint probability.
Its supported interpretation is the relative ordering of conformations and ensembles with respect to the two learned endpoint models.

\subsection{Regional decomposition reveals distinct extracellular and intracellular contributions to endpoint preference}
\label{sec:regional_endpoint}

The explicit residue-wise structure of the Potts Hamiltonian makes it possible to ask not only whether a conformation is preferentially supported by the inactive or active endpoint model, but also where in the receptor this preference originates.
We therefore decompose $\Delta E_{\mathrm{INzma-FANG}}$ into contributions associated with two functionally distinct regions: an extracellular region containing the orthosteric binding pocket and extracellular loops, and an intracellular region containing the cytoplasmic portions of the transmembrane helices and the intracellular loops.

\begin{figure}[ht]
\captionsetup{font={small,stretch=1.0}}
\vspace{-40pt}
\centering
\includegraphics[width=0.8\textwidth]{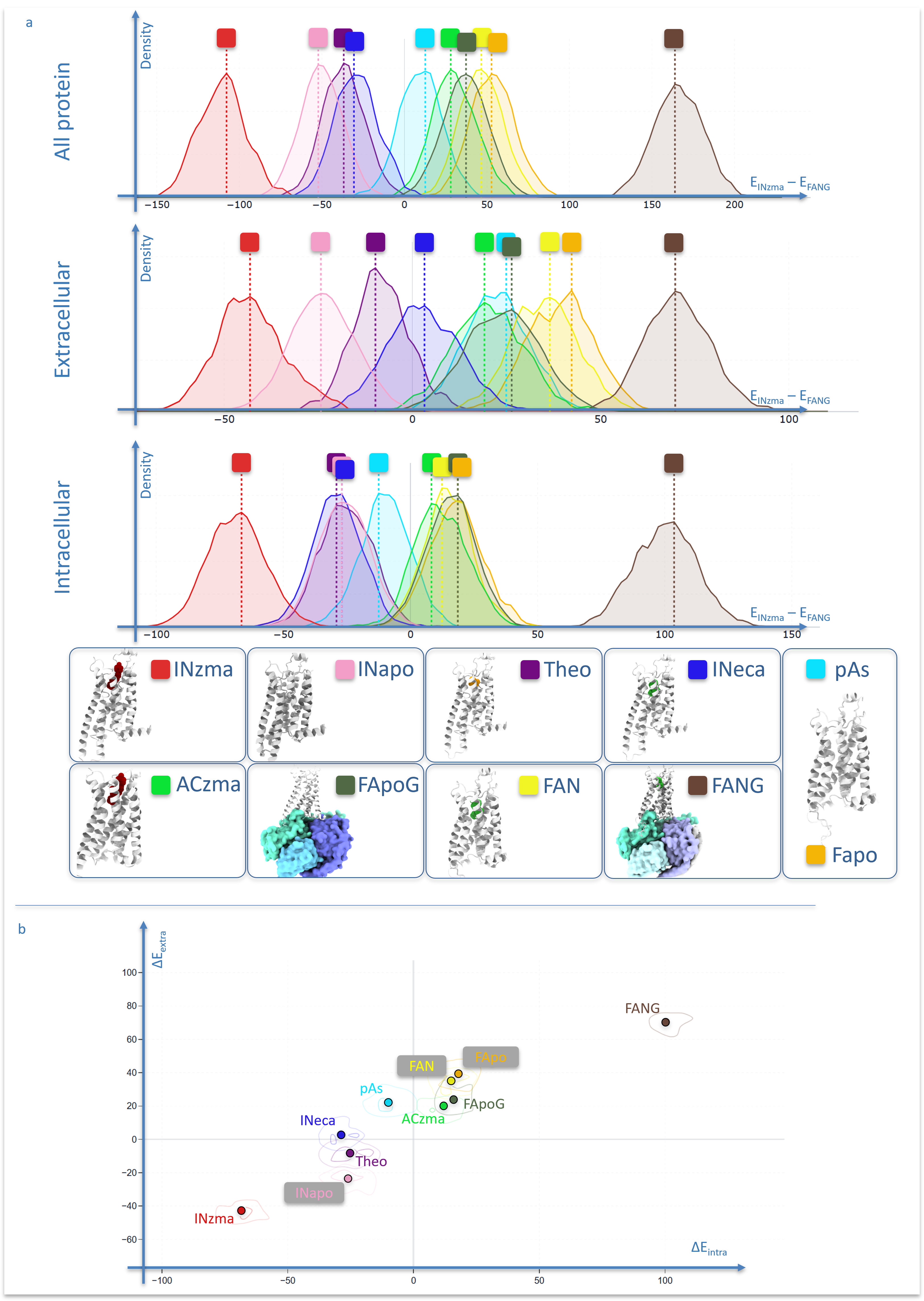}

\caption{\textbf{Replica-exchange sampling preserves global and regional endpoint organization across \AtwoA{} ensembles.}
\textbf{a,} Distributions of the endpoint preference score $\Delta E_{\mathrm{INzma-FANG}}$ for configurations generated independently by replica-exchange Gibbs sampling from each state-specific \PHASE{} Hamiltonian. Scores are evaluated using the complete receptor (top), the extracellular region containing the orthosteric pocket and extracellular loops (middle), and the intracellular region containing the cytoplasmic portions of the transmembrane helices and intracellular loops (bottom). Increasing values indicate greater relative preference for the FANG endpoint Hamiltonian. \textbf{b,} Joint representation of the extracellular and intracellular endpoint scores for the same generated configurations.
For visualization, the approximate position of each distribution is summarized by its modal value, and contours represent its distribution in the two-dimensional regional score space. INzma and FANG anchor opposite ends of the inactive--active landscape. Corresponding score distributions evaluated directly on the source MD ensembles are shown in Supplementary Fig.~\ref{fig:supp_md_vs_sampling_energies}.}


\label{fig:regional_endpoint}
\end{figure}

For the same replica-exchange-generated configurations, the complete-receptor endpoint score is shown in Fig.~\ref{fig:regional_endpoint}a (top), together with the extracellular (middle) and intracellular (bottom) regional scores.
The decomposition shows that the global inactive--active coordinate is not encoded uniformly throughout the receptor.
Much of the separation among ligand-conditioned ensembles is already apparent in the extracellular score, consistent with the established sensitivity of the \AtwoA{} orthosteric pocket and extracellular network to ligand efficacy and with propagation of these changes through the receptor \cite{Ye2016Nature,Lebon2011Nature,DAmore2024Chem}.
The intracellular score, however, reorganizes several relationships between ensembles, indicating that the ligand-facing and transducer-facing regions contain partially distinct information about receptor state.

The sampled FApo--pAs ensembles illustrate a global conformational relaxation.
These ensembles correspond to consecutive portions of the same apo simulation and therefore differ neither in ligand occupancy nor in effector coupling.
Nevertheless, the transition from the initially active-like FApo segment to the subsequently stabilized pAs ensemble shifts both extracellular and intracellular score distributions.
The spontaneous relaxation toward pAs is therefore not confined to the cytoplasmic receptor but involves a coordinated reorganization across the protein, consistent with the distributed microswitch changes associated with the pseudo-active state \cite{DAmore2024Chem}.

A different relationship is preserved between the sampled pAs and FApoG ensembles.
Both ensembles are apo, and their extracellular score distributions substantially overlap, indicating a similar degree of active-endpoint preference in the ligand-facing region.
Their intracellular distributions are comparatively more displaced, with the mini-Gs-coupled FApoG ensemble shifted toward the active endpoint.
This is consistent with the established role of G-protein or mini-G-protein coupling in stabilizing the intracellular active architecture of \AtwoA, including the cytoplasmic transmembrane-helical rearrangements required for transducer engagement \cite{Carpenter2016Nature,Lee2019Structure,DAmore2024Chem}.
The comparison is particularly informative because it shows that similarity in the extracellular receptor does not necessarily imply an equivalent intracellular conformational preference.

The complementary nature of the two regional signals becomes clearer when each ensemble is represented jointly by its extracellular and intracellular endpoint scores (Fig.~\ref{fig:regional_endpoint}b).
The stabilized endpoints occupy opposite corners of this two-dimensional space, whereas the remaining ensembles trace different combinations of extracellular and intracellular preference.
INapo, Theo and INeca progressively depart from the inactive reference primarily through changes associated with the ligand-facing side of the receptor, while the active-side ensembles additionally separate according to their intracellular organization.
FApo and pAs illustrate coordinated motion along both components, whereas comparisons such as pAs and FApoG reveal that similar extracellular preferences can coexist with different intracellular states.

The regional decomposition therefore adds information that is hidden when the Hamiltonian is reduced to a single global endpoint score.
Two ensembles can have similar global or extracellular preferences while differing in the region directly involved in transducer coupling.
\PHASE{} can consequently be used not only to rank conformations along an inactive--active coordinate, but also to construct spatially resolved statistical fingerprints of receptor activation.
Such fingerprints provide a natural basis for future analyses of functional selectivity, where ligands producing comparable orthosteric activation may nevertheless stabilize different intracellular conformations associated with distinct signalling partners.

\subsection{Cluster-conditioned backmapping closes the generative cycle}
\label{sec:backmapping_res}
Reduced-space samples become structurally useful only if they can be realized as atomistic conformations. We therefore modify and fine-tune a protein folding model, SimpleFold \cite{Wang2025SimpleFold}, to receive the \AtwoA{} sequence together with one cluster identity per residue (see \S \ref{subsec:cluster_conditioned_backmapping}).

\noindent We fine-tune on subsampled frames from the \AtwoA{} INzma, INapo, Theo, and FANG trajectories and evaluate the backmapping model in two complementary ways on a held-out test set comprising frames from MD simulations of the same ensembles.

\begin{figure}[ht]
\captionsetup{font={small,stretch=1.0}}
\centering
\includegraphics[width=\textwidth]{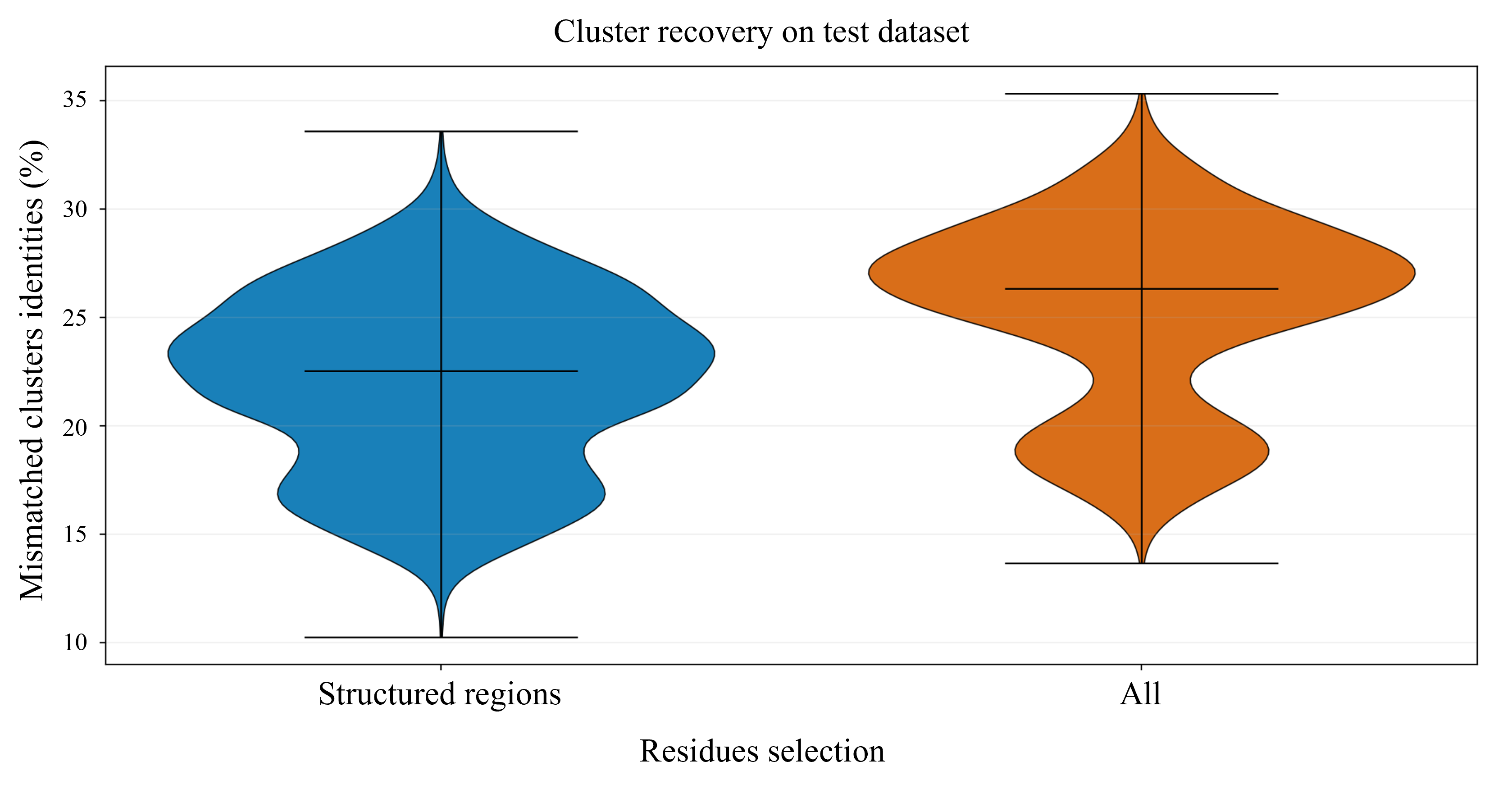}
\caption{Held-out cluster recovery for backmapped Potts Hamiltonian samples after re-projection for the different residues selection.}
\label{fig:backmapping_clu_mism_res}
\end{figure}

\begin{figure}[ht]
\captionsetup{font={small,stretch=1.0}}
\centering
\includegraphics[width=\textwidth]{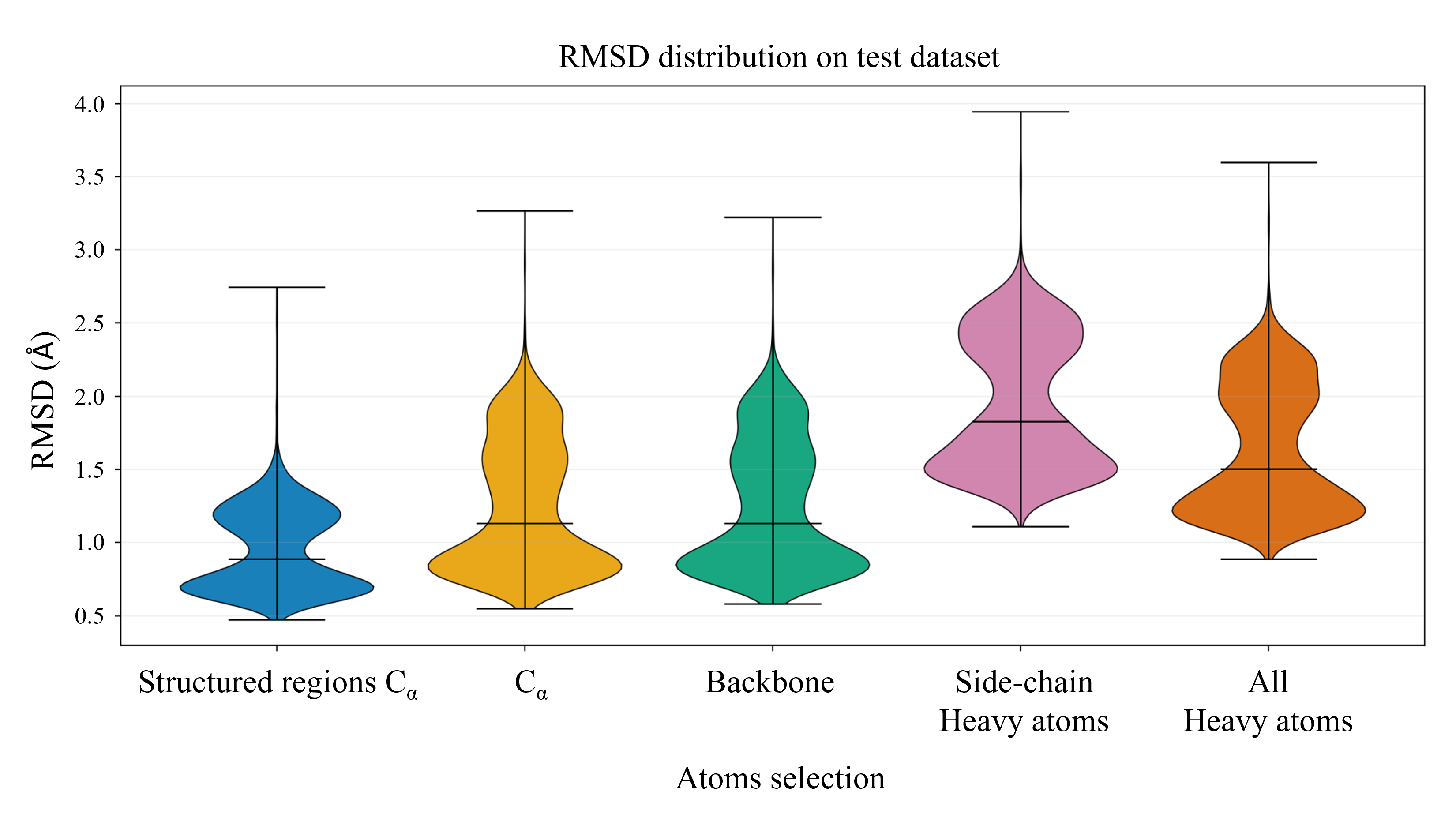}
\caption{Held-out atomistic reconstruction. RMSDs evaluated on C$\alpha$, backbone, side-chain heavy-atoms, all-heavy-atoms and structured regions $C_\alpha$.}
\label{fig:backmapping_rmsd_res}
\end{figure}

Re-coarse-graining each prediction assesses whether the sampled all-atom structure is able to reproduce the input microstate vector. The percentage of mismatched clusters identities is measured over the complete receptor and over the selection of residues that define its structured regions (see \S \ref{subsec:backmapping_evaluation}). Coordinate reconstruction is evaluated by aligned RMSD for C$\alpha$, backbone, side-chain heavy-atoms, all-heavy-atoms and C$\alpha$ of the structured regions. The resulting distributions exhibit an average cluster-identity mismatch of 22.1\% $\pm$ 4.1 in structured regions and 25.3\% $\pm$ 4.3 across all residues (Fig. \ref{fig:backmapping_clu_mism_res}), while the corresponding RMSD results are reported in Table \ref{tab:rmsd_aggrg} and Fig. \ref{fig:backmapping_rmsd_res}. We achieve sub-angstrom accuracy on the $C_\alpha$ of the structured regions of the receptor. All metrics reported are aggregated over the different conformational ensembles.



\begin{table*}[t]
\centering
\small
\begin{tabular}{l|l}
\toprule
Residue selection critera & RMSD  ($\mathring{A})$ \\
\midrule
All-heavy-atoms & 1.63 $\pm$ 0.43 \\
$C_\alpha$ & 1.25 $\pm$ 0.44 \\
Backbone & 1.25 $\pm$ 0.44 \\
Side-chain heavy-atoms & 1.94 $\pm$ 0.45 \\
Structured regions $C_\alpha$ & 0.95 $\pm$ 0.28 \\
\bottomrule
\end{tabular}
\caption{RMSD errors between reference and backmapped structures for the different residues selection. All metrics reported are aggregated over the different conformational ensembles.}
\label{tab:rmsd_aggrg}
\end{table*}

\begin{figure}[ht]
    \captionsetup{font={small,stretch=1.0}}
    \centering

    \begin{subfigure}{0.32\textwidth}
        \centering
        \includegraphics[width=\textwidth]{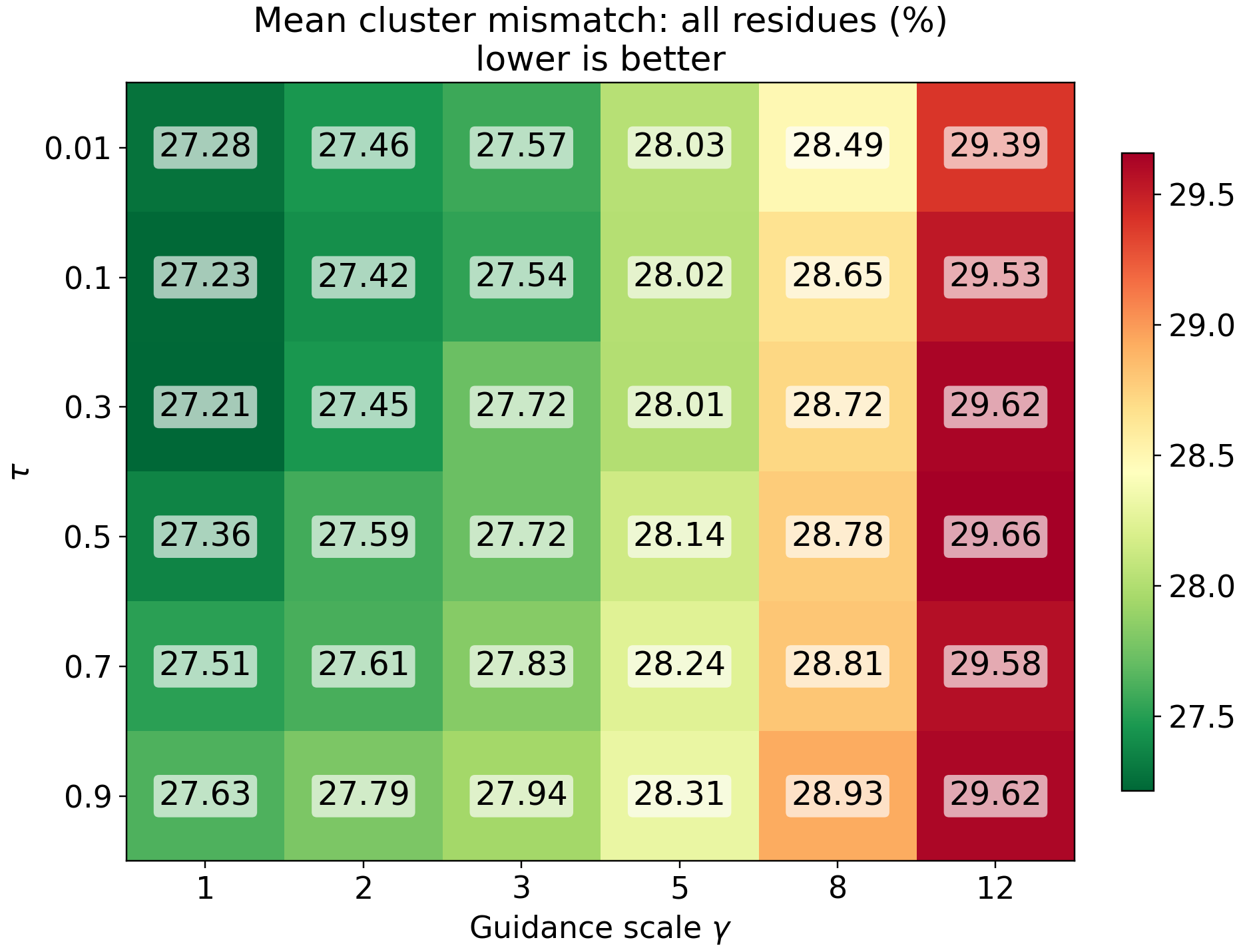}
        \caption{}
        \label{fig:image1}
    \end{subfigure}
    \hfill
    \begin{subfigure}{0.32\textwidth}
        \centering
        \includegraphics[width=\textwidth]{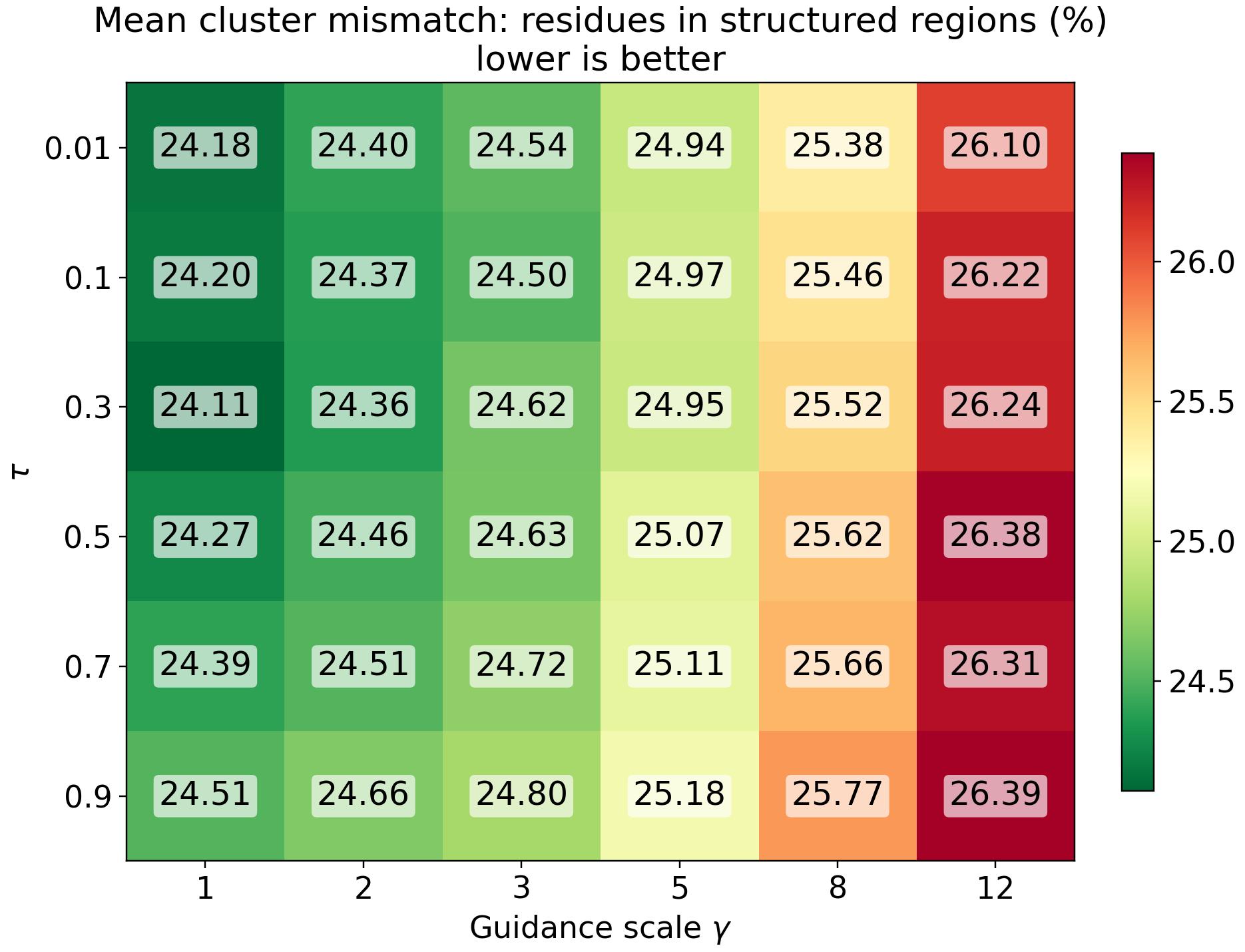}
        \caption{}
        \label{fig:image2}
    \end{subfigure}
    \hfill
    \begin{subfigure}{0.32\textwidth}
        \centering
        \includegraphics[width=\textwidth]{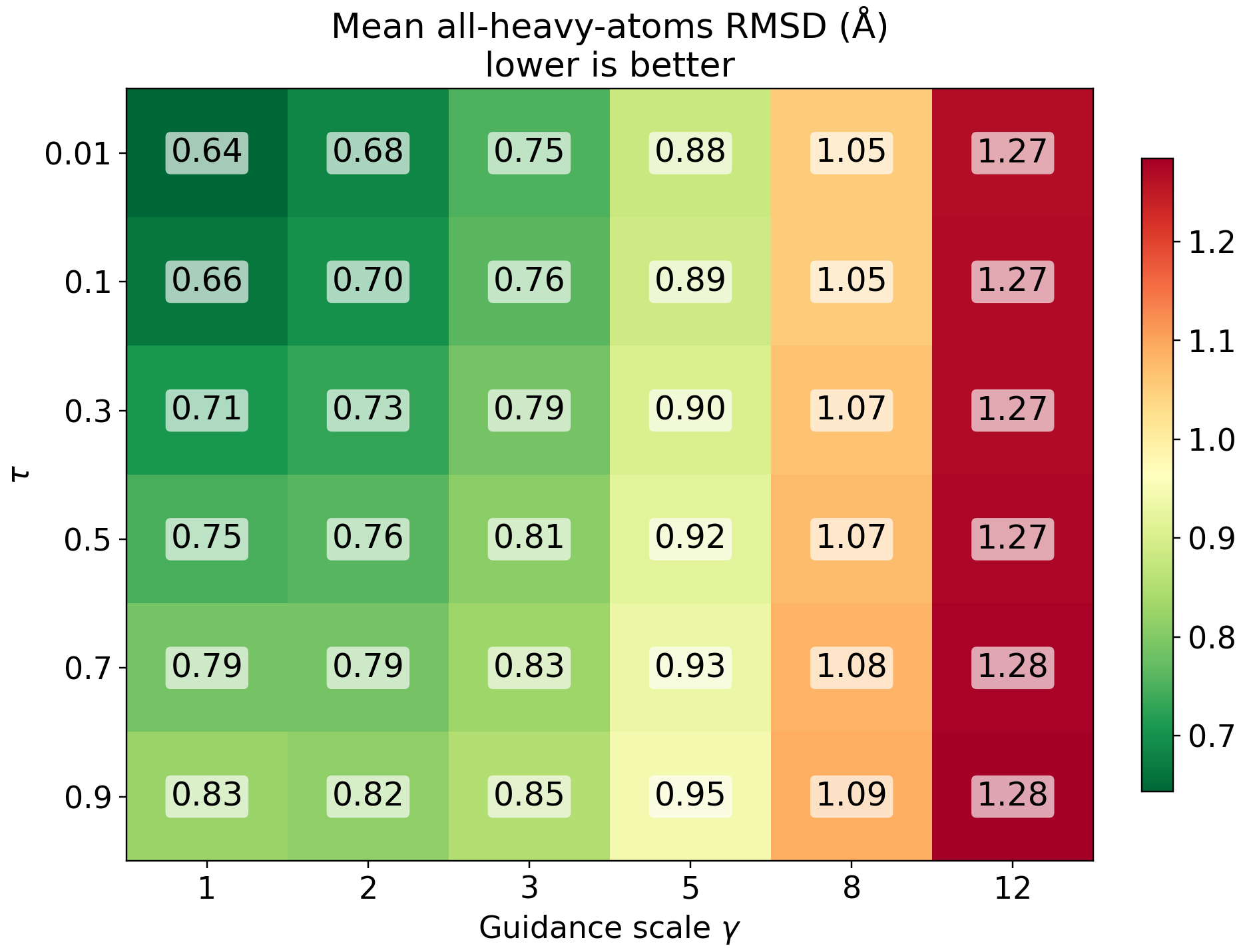}
        \caption{}
        \label{fig:image3}
    \end{subfigure}

    \vspace{0.5em}

    \begin{subfigure}{0.32\textwidth}
        \centering
        \includegraphics[width=\textwidth]{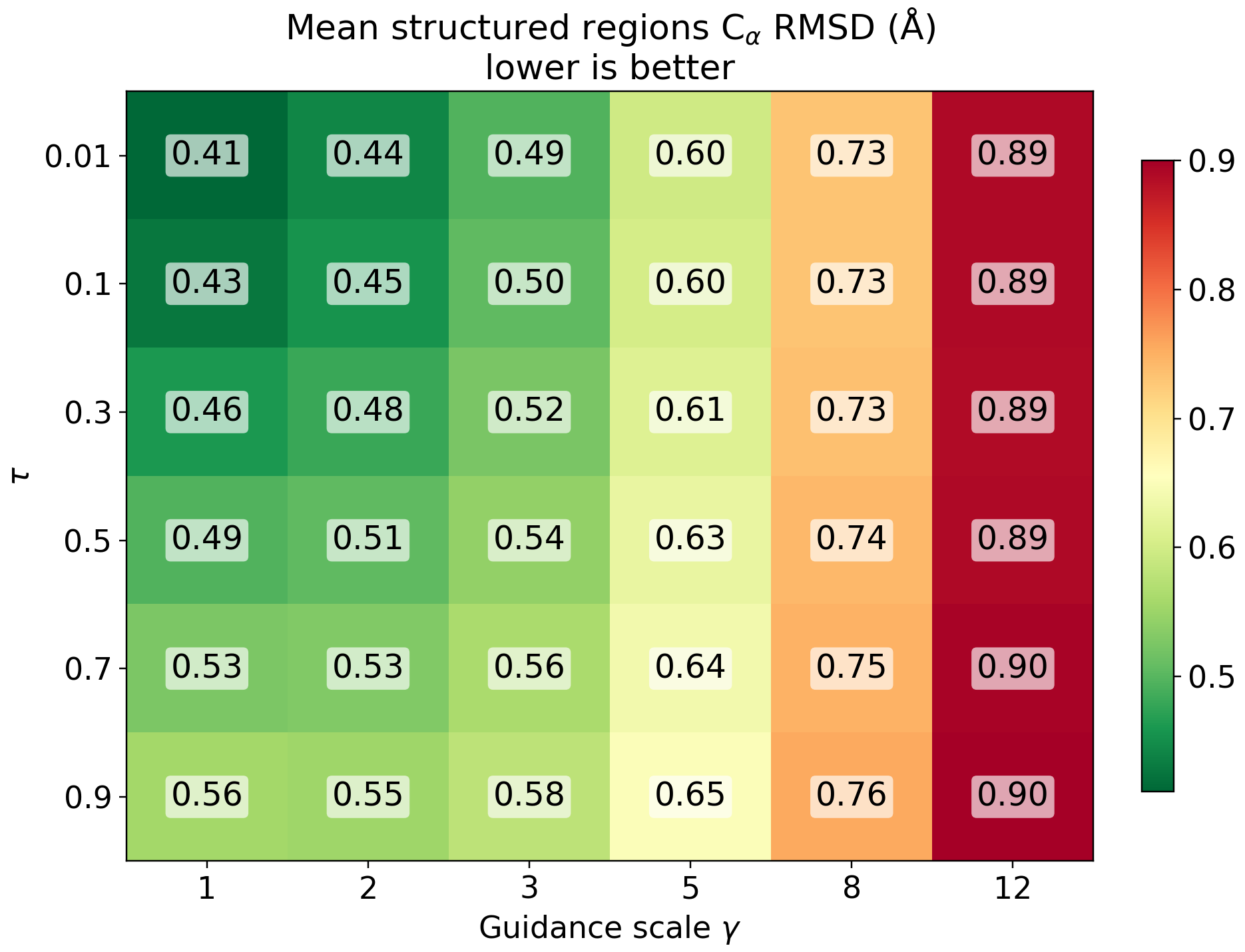}
        \caption{}
        \label{fig:image4}
    \end{subfigure}
    \hfill
    \begin{subfigure}{0.32\textwidth}
        \centering
        \includegraphics[width=\textwidth]{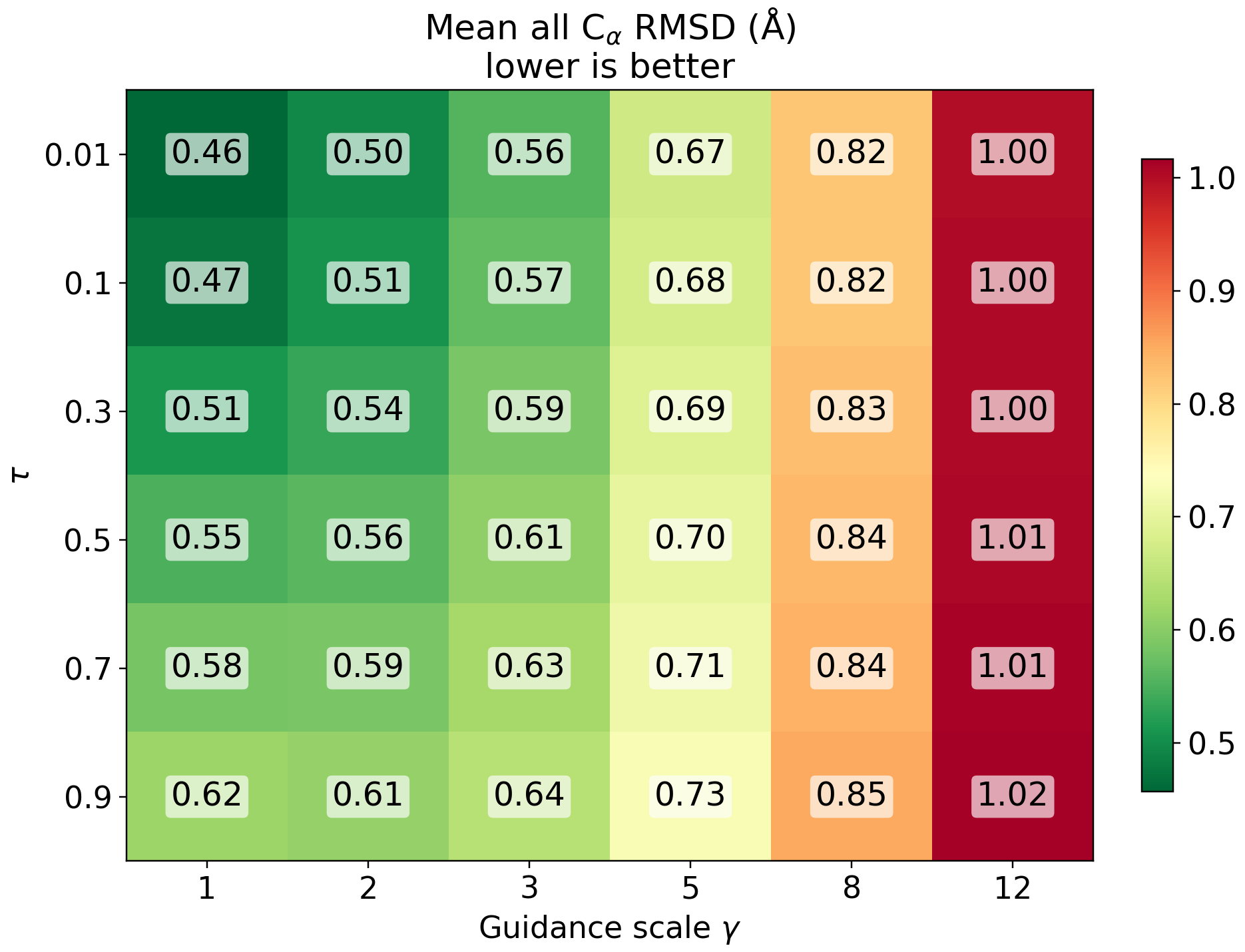}
        \caption{}
        \label{fig:image5}
    \end{subfigure}
    \hfill
    \begin{subfigure}{0.32\textwidth}
        \centering
        \includegraphics[width=\textwidth]{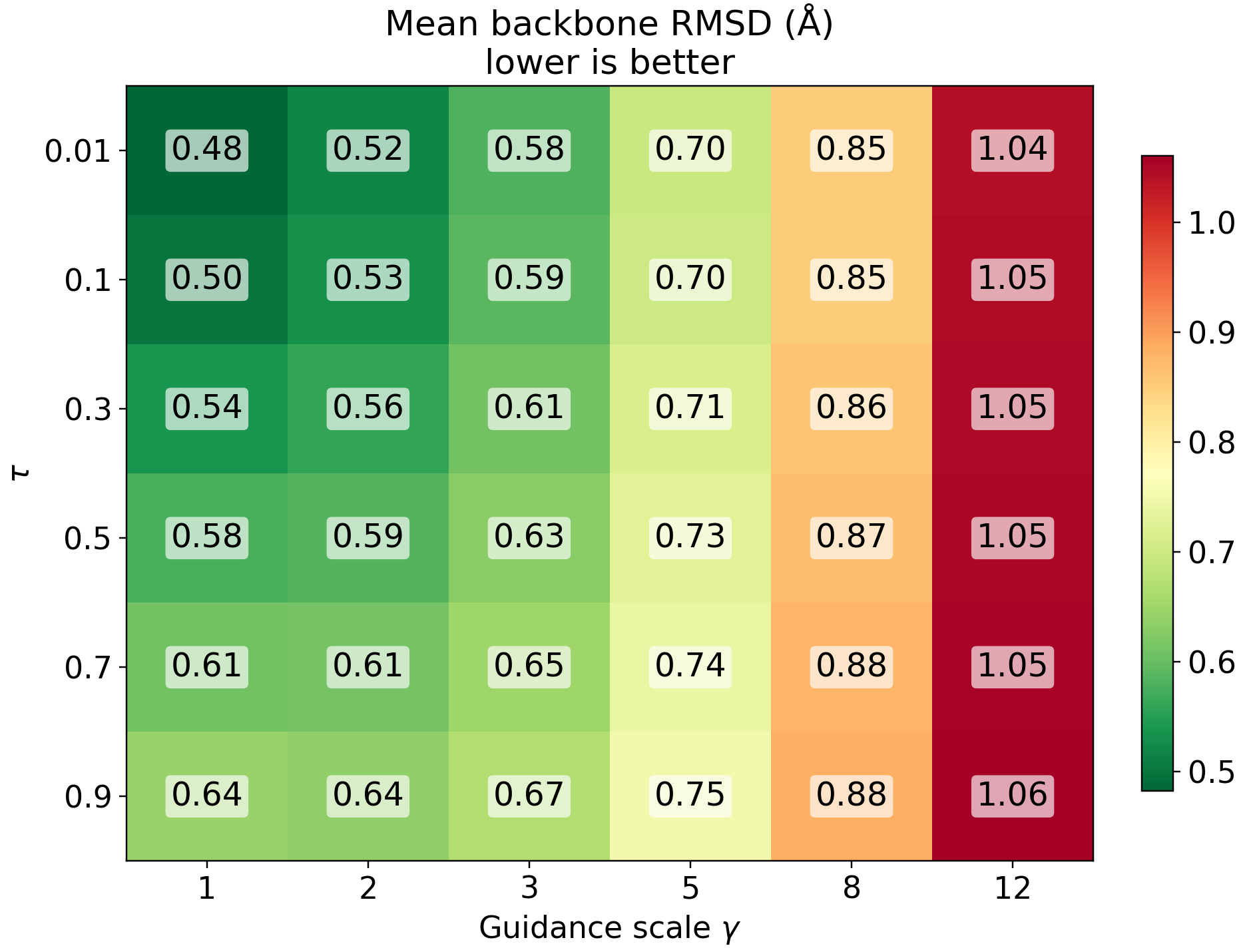}
        \caption{}
        \label{fig:image6}
    \end{subfigure}
    \caption{Results of the grid search of $\tau$ (the scale of stochasticity at sampling time) and the guidance scale $\gamma$. The most frequent best combination of parameters is $\tau=0.01$ and $\gamma=1.0$. The sweep has been executed by backmapping a random subsample of 100 structures per conformational ensemble in the training dataset, the values shown are averages across the whole subsample. (a) \% of mismatched clusters across all residues (b) \% of mismatched clusters across all residues of structured regions (c) RMSD computed over all-heavy-atoms (d) RMSD computed over $C_{\alpha}$ of the residues of structured regions (e) RMSD on $C_\alpha$ only (f) RMSD on backbone.}
    \label{fig:backmapping_sweep}
\end{figure}

\noindent The results provided are obtained after executing a grid search for the two main inference hyper-parameters $\tau$ (the scale of stochasticity at sampling time) and the guidance scale $\gamma$ (see \S \ref{subsec:cluster_conditioned_backmapping}). The sweep is executed by backmapping a random subsample of 100 structures per conformational ensemble over the training dataset. In Fig. \ref{fig:backmapping_sweep} we show that best combination of parameters is $\tau=0.01$ and $\gamma=1.0$.

\begin{figure}[ht]
    \captionsetup{font={small,stretch=1.0}}
    \centering
    \begin{subfigure}{0.35\textwidth}
        \centering
        \includegraphics[width=\textwidth]{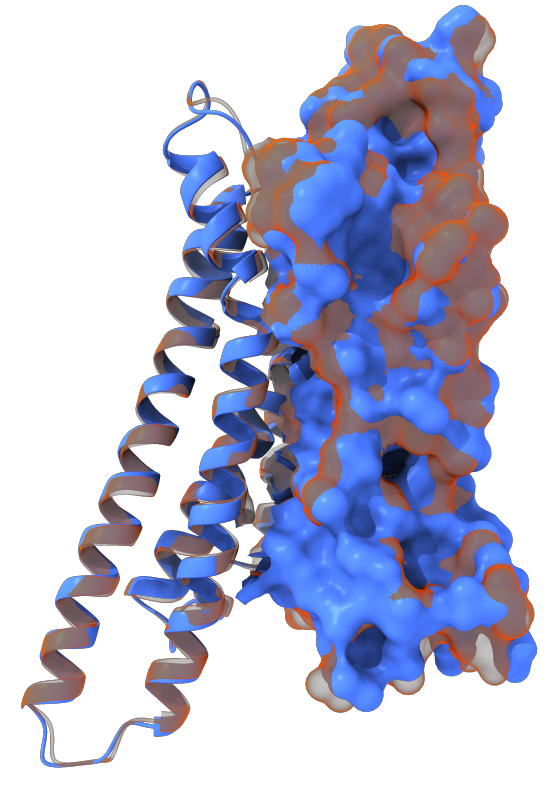}
        \caption{}
        \label{fig:image1}
    \end{subfigure}
    \hspace{3.5em}
    \begin{subfigure}{0.35\textwidth}
        \centering
        \includegraphics[width=\textwidth]{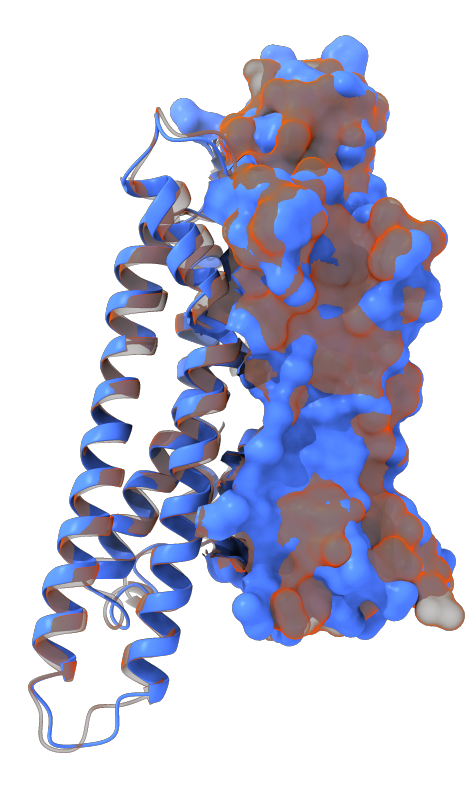}
        \caption{}
        \label{fig:image2}
    \end{subfigure}    
    \vspace{0.5em}

    \begin{subfigure}{0.35\textwidth}
        \centering
        \includegraphics[width=\textwidth]{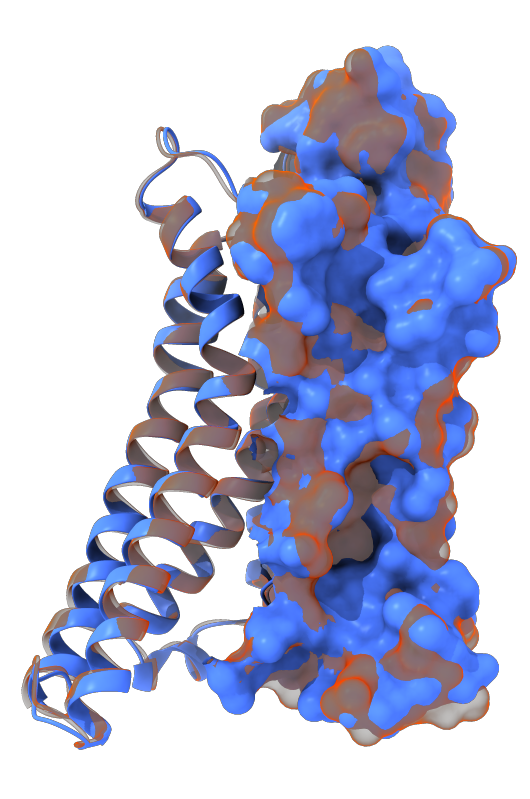}
        \caption{}
        \label{fig:image4}
    \end{subfigure}
    \hspace{3.5em}
    \begin{subfigure}{0.35\textwidth}
        \centering
        \includegraphics[width=\textwidth]{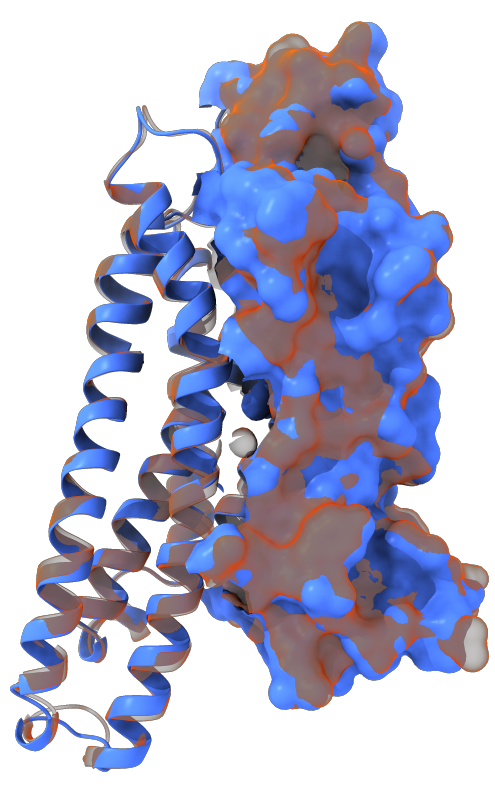}
        \caption{}
        \label{fig:image7}
    \end{subfigure}
    \caption{Backmapped structures $\hat{\textbf{R}}$ in blue overlapped with their associated ground truth $\textbf{R*}$ in red.
    The reference structures are selected from the test set.
    (a) INapo, all-heavy-atoms RMSD: 0.98 (b) INzma, all-heavy-atoms RMSD: 1.25 (c) FANG, all-heavy-atoms RMSD: 0.89 (d) Theo, all-heavy-atoms RMSD: 0.96}
    \label{fig:og_vs_backmapped}
\end{figure}

We then backmap configurations sampled directly from the Potts Hamiltonian. These states do not have a unique atomistic target, so the RMSD to an original frame is undefined.
\noindent This experiment closes the \PHASE{} cycle: the Potts Hamiltonian model generates statistically weighted coarse-grained configurations, and the conditioned structure backmapping model realizes them while preserving their defining local states. For a set of 1,000 generated configurations from a Potts Hamiltonian fitted onto INzma, we obtain 22.2\% $\pm$ 2.3 mismatched clusters on the structured regions and 26.4\% $\pm$ 2.2 on the whole receptor. 
Fig. \ref{fig:og_vs_backmapped} visualizes the backmapped structures $\hat{\textbf{R}}$ overlapped with their associated ground truth $\textbf{R*}$ for each ensemble on which we have trained the backmapping model.

\section{Discussion}

\PHASE{} provides a system-specific route from atomistic protein ensembles to an explicit, interpretable and generative statistical model. Molecular dynamics supplies the conformational information, while \PHASE{} compresses it into a shared residue-state representation and a sparse Potts Hamiltonian whose variables and interactions can be inspected, compared across conditions, sampled and mapped back to atomistic structures. The procedure is protein-general, but the microstate basis, interaction graph and statistical parameters are learned specifically for the molecular system under study.

A central practical advantage of PHASE is that the fitted Hamiltonian defines an explicit generative distribution over the conformational ensemble, from which new combinations of residue microstates can be sampled. In this work, we use Gibbs and replica-exchange sampling, but the representation is not tied to a particular sampling algorithm: alternative Monte Carlo strategies can be employed depending on the desired trade-off between simplicity, computational efficiency, mixing, and coverage of the learned distribution \cite{MonteCarlo-NewmanBarkema1999,MonteCarlo-LandauBinder2014}.
For A2A, we show that the generated configurations reproduce the residue-wise and pairwise statistics of the source ensembles and preserve their higher-level organization along the inactive–active preference coordinate. Combined with cluster-conditioned all-atom backmapping, this makes it possible to repeatedly draw atomistically realizable conformations from the learned ensemble. 
An appealing extension would be to fit the Hamiltonian from enhanced-sampling-biased ensembles, by incorporating importance weights into the pseudolikelihood objective, thereby correcting the contribution of individual configurations toward a desired target distribution. This could substantially broaden the range of simulation data usable for PHASE, allowing the model to exploit trajectories designed to explore a larger fraction of conformational space while retaining a statistically defined target ensemble.

A second interesting use of PHASE follows from representing different ensembles in the same residue-state space. Once the microstate definitions and contact graph are fixed, independently fitted Hamiltonians differ only in their statistical parameters and can therefore be compared after gauge normalization. In \AtwoA{}, Hamiltonians fitted on the inactive INzma and active FANG are used as reference endpoints. Their difference defines the endpoint preference score: each configuration is evaluated under both Hamiltonians, and the difference between the resulting energies quantifies its relative compatibility with the two endpoints, providing a one-dimensional ordering between the inactive and active references.
Simulations performed with different ligands, starting conformations and effector conditions distribute systematically along this score, despite none of this biochemical information being supplied to the models. 
The score remains a relative statistical compatibility measure and not a physical inactive--active free-energy difference.

The explicit residue-wise structure of the Hamiltonian also makes these differences spatially interpretable. Extracellular and intracellular decompositions reveal that the global activation coordinate contains distinct regional contributions. Ligand-dependent differences are strongly represented around the orthosteric and extracellular region, whereas mini-Gs coupling produces additional separation in the intracellular receptor. In particular, pAs and FApoG show similar extracellular preferences but differ more clearly in their intracellular signatures, consistent with stabilization of the transducer-facing active architecture by mini-Gs \cite{Carpenter2016Nature,Lee2019Structure,DAmore2024Chem}. This type of regional statistical fingerprint could be useful for comparing ligand efficacy, allosteric modulation or different intracellular partners. In receptors, it also provides a natural framework for future studies of functional selectivity.

For a new protein, \PHASE{} is particularly useful when multiple atomistic ensembles must be compared under different conditions, such as ligand binding, mutation, allosteric perturbation or association with different molecular partners.
\PHASE{} projects these trajectories in a common residue-state space and summarizes each ensemble with an explicit Hamiltonian.
Differences can then be traced to residue fields and pair interactions, candidate conformations can be scored against one or more reference ensembles, and the learned distributions can be sampled to generate new residue-state combinations for all-atom reconstruction.
\PHASE{} therefore provides a reusable statistical representation of the ensemble that can be interrogated after the original simulations have been generated.


The main limitations follow directly from this construction. \PHASE{} cannot be expected to recover conformational basins that are absent from the data used to define and fit the model, and therefore inherits sampling and force-field biases of the underlying simulations. Likewise, the interpretation of endpoint-based comparisons depends on the reference Hamiltonians providing representative anchors for the coordinate of interest (in the case of A2A, the coordinate has a functional interpretation, spanning from inactive to active). If an ensemble selected as endpoint does not represent an extremal within the range of variation of such coordinate, interpretation of the ranking along the resulting preference coordinate can become incomplete or misleading.
Its torsional discretization and pairwise Hamiltonian further assume that the statistics of interest can be represented sufficiently well through residue states and sparse pair interactions; marginal reconstruction, non-edge pair statistics and held-out configuration tests provide direct diagnostics of this approximation.
Finally, Monte Carlo sampling target the learned statistical distribution rather than molecular dynamics or kinetics, so generated sequences of microstates should not be interpreted as physical transition trajectories.

Finally, the discrete PHASE Hamiltonian additionally admits an exact one-hot QUBO representation, enabling exploration with binary annealing algorithms (see Methods \ref{subsec:qubo}).
Our classical simulated-annealing baseline shows that the resulting exploration depends on the enforcement of the one-hot constraints and accesses different regions of the PHASE energy landscape than equilibrium Gibbs sampling.
The QUBO formulation therefore provides an additional representation for constrained optimization and energy-landscape exploration.
Developing annealing protocols capable of reliably identifying diverse low-energy conformations of the learned Potts landscape remains an interesting direction for future work, both on specialized classical and quantum annealing hardware.

Overall, \PHASE{} bridges atomistic ensemble simulation and explicit statistical modelling. Its main contribution is not simply the generation of additional structures, but a compact Hamiltonian in which conformational statistics become local, interpretable and manipulable while retaining a route back to atomistic coordinates.

\section{Methods}
\label{sec:methods}

\subsection{Atomistic trajectories and notation}
\label{subsec:trajectory_notation}

Consider a protein with a fixed amino-acid sequence
\begin{equation}
    \mathbf{a}=(a_1,\ldots,a_N),
\end{equation}
where $N$ is the number of modeled residues and $a_r$ is the residue type at sequence position $r$. For the \AtwoA{} receptor analyzed here, $N=300$.

Let $A$ denote the number of retained protein atoms, and let
\begin{equation}
    \mathcal{A}_r\subseteq\{1,\ldots,A\}
\end{equation}
be the set of atom indices belonging to residue $r$. The sets $\mathcal{A}_1,\ldots,\mathcal{A}_N$ define the atom-to-residue mapping and are fixed across all trajectories.

A molecular-dynamics dataset contains trajectories indexed by $q\in\mathcal{Q}$. Trajectory $q$ contains $F_q$ stored frames. The protein coordinates in frame $f$ are represented by
\begin{equation}
    \mathbf{R}^{(q,f)}
    =
    \left(
    \mathbf{r}^{(q,f)}_1,\ldots,
    \mathbf{r}^{(q,f)}_A
    \right)
    \in\mathbb{R}^{A\times 3},
    \qquad
    f\in\{1,\ldots,F_q\},
    \label{eq:atomistic_frame}
\end{equation}
where $\mathbf{r}^{(q,f)}_i\in\mathbb{R}^3$ is the Cartesian position of atom $i$. Only protein coordinates enter the \PHASE{} representation. Ligand, membrane, solvent, ions and, when present, mini-Gs coordinates are excluded from $\mathbf{R}^{(q,f)}$.

The \AtwoA{} dataset comprises eight underlying atomistic MD trajectories (Table~\ref{tab:a2a_systems}).
INzma contains 1.26~$\mu$s of simulation and Theo contains 1.0~$\mu$s, whereas INapo, INeca, FApo, FApoG, FAN and FANG contain 5.0~$\mu$s each.
Frames are analyzed at 100~ps intervals, yielding 12,600 frames for INzma, 10,000 frames for Theo and 50,000 frames for each 5~$\mu$s trajectory.

The active-starting apo trajectory requires an additional temporal partitioning.
During this trajectory, the receptor leaves its initial active-like region and stabilizes in the pseudo-active state identified in the underlying \AtwoA{} activation analysis.
We therefore divide the trajectory at 1.5~$\mu$s.
Frames from the first 1.5~$\mu$s define the FApo ensemble, whereas frames from 1.5 to 5.0~$\mu$s define the pAs ensemble.
Accordingly, FApo contains 15,000 analyzed frames and pAs contains 35,000 frames.
Although these two ensembles are treated separately in the subsequent statistical analyses, they originate from the same continuous MD trajectory.

\subsection{Residue torsional descriptors}
\label{subsec:torsional_descriptors}

Each atomistic frame is reduced to one local angular descriptor per residue. Note that two residues that share the same residue type (e.g. alanine) but different residue index will have different and unique descriptors.
For four ordered atoms with positions $\mathbf{u}_1,\ldots,\mathbf{u}_4$, let
\begin{equation}
    \operatorname{dih}(\mathbf{u}_1,\mathbf{u}_2,\mathbf{u}_3,\mathbf{u}_4)
    \in[-\pi,\pi)
\end{equation}
denote the signed dihedral angle defined by the two planes $(\mathbf{u}_1,\mathbf{u}_2,\mathbf{u}_3)$ and $(\mathbf{u}_2,\mathbf{u}_3,\mathbf{u}_4)$. The backbone torsions of residue $r$ in frame $(q,f)$ are
\begin{align}
    \phi_r^{(q,f)}
    &=
    \operatorname{dih}
    \left(
    \mathbf{r}_{\mathrm{C}_{r-1}}^{(q,f)},
    \mathbf{r}_{\mathrm{N}_{r}}^{(q,f)},
    \mathbf{r}_{\mathrm{C}\alpha_{r}}^{(q,f)},
    \mathbf{r}_{\mathrm{C}_{r}}^{(q,f)}
    \right),\\
    \psi_r^{(q,f)}
    &=
    \operatorname{dih}
    \left(
    \mathbf{r}_{\mathrm{N}_{r}}^{(q,f)},
    \mathbf{r}_{\mathrm{C}\alpha_{r}}^{(q,f)},
    \mathbf{r}_{\mathrm{C}_{r}}^{(q,f)},
    \mathbf{r}_{\mathrm{N}_{r+1}}^{(q,f)}
    \right),\\
    \omega_r^{(q,f)}
    &=
    \operatorname{dih}
    \left(
    \mathbf{r}_{\mathrm{C}\alpha_{r-1}}^{(q,f)},
    \mathbf{r}_{\mathrm{C}_{r-1}}^{(q,f)},
    \mathbf{r}_{\mathrm{N}_{r}}^{(q,f)},
    \mathbf{r}_{\mathrm{C}\alpha_{r}}^{(q,f)}
    \right).
    \label{eq:backbone_torsions}
\end{align}
The side-chain torsions $\chi_{1,r}^{(q,f)}$ and $\chi_{2,r}^{(q,f)}$ are defined by the standard residue-specific atom quadruplets when those degrees of freedom exist. The complete fixed-format descriptor is
\begin{equation}
    \boldsymbol{\theta}_r^{(q,f)}
    =
    \left(
    \phi_r^{(q,f)},
    \psi_r^{(q,f)},
    \omega_r^{(q,f)},
    \chi_{1,r}^{(q,f)},
    \chi_{2,r}^{(q,f)}
    \right).
    \label{eq:torsional_descriptor}
\end{equation}
Not every component is defined for every residue. Terminal backbone angles and absent side-chain torsions (e.g., $\chi_1$ and $\chi_2$ for alanine) are excluded from the effective clustering coordinates. Let $\mathcal{I}_r\subseteq\{\phi,\psi,\omega,\chi_1,\chi_2\}$ denote the informative angular components retained for residue $r$, and let $d_r=|\mathcal{I}_r|$.

All angular variables are treated periodically. For two angles $\alpha$ and $\gamma$, their wrapped difference is
\begin{equation}
    \delta_{2\pi}(\alpha,\gamma)
    =
    \operatorname{atan2}
    \left[
    \sin(\alpha-\gamma),
    \cos(\alpha-\gamma)
    \right].
    \label{eq:wrapped_angular_difference}
\end{equation}
Accordingly, the residue descriptor belongs to a $d_r$-dimensional torus.

For selected residue types with chemically equivalent $\chi_2$ conformations separated by $\pi$, the second side-chain angle is folded before clustering and before projecting new frames:
\begin{equation}
    \widetilde{\chi}_{2,r}^{(q,f)}
    =
    \left(2\chi_{2,r}^{(q,f)}\right)\bmod 2\pi.
    \label{eq:symmetric_chi2_folding}
\end{equation}
This mapping identifies $\chi_2$ and $\chi_2+\pi$ as the same periodic coordinate. In the present implementation it is applied to phenylalanine, tyrosine and aspartate.

Let
\begin{equation}
    \mathbf{u}_r^{(q,f)}
    =
    \mathcal{P}_r\!\left(\boldsymbol{\theta}_r^{(q,f)}\right)
    \in\mathbb{T}^{d_r}
    \label{eq:preprocessed_descriptor}
\end{equation}
denote the residue-specific preprocessing map that selects the components in $\mathcal{I}_r$, maps them to $[0,2\pi)$ and applies the symmetry correction in Eq.~\eqref{eq:symmetric_chi2_folding} where required.

\subsection{Shared residue-wise microstate basis}
\label{subsec:shared_microstate_basis}

\PHASE{} represents each residue $r$ by a discrete microstate label derived from clustering in the space of local torsional conformations $\Theta_r$. The key idea is that this microstate basis is learned \emph{once} from a common pool of structures and then shared across all simulations considered in the study. In this way, inactive, active, and intermediate ensembles can be compared in a common residue-level state space: different simulations may populate the microstates with different frequencies, but the underlying microstate definition remains fixed.

Let $\mathcal Q_{\mathrm{basis}}$ denote the complete set of ten available \AtwoA{} trajectories.
All trajectories in $\mathcal Q_{\mathrm{basis}}$ are pooled to construct the shared residue-wise microstate basis used throughout the analysis.
For residue $r$, frame $f$ of trajectory $q$ is represented by a vector of local torsional variables,
\begin{equation}
\boldsymbol{\theta}_{r}^{(q,f)}
=
\bigl(\theta_{r,1}^{(q,f)},\ldots,\theta_{r,d_r}^{(q,f)}\bigr) \in \Theta_r\,,
\label{eq:residue_torsion_vector}
\end{equation}
where the dimension $d_r$ depends on the residue type and on the torsional degrees of freedom retained for that residue. These include the relevant backbone and side-chain dihedral angles. Since torsions are periodic, the representation is constructed so that angular proximity is treated correctly across periodic boundaries. In particular, symmetric side-chain torsions such as $\chi_2$ are wrapped into their symmetry-reduced domain before clustering.

For each residue $r$, the density of points from the pooled set
\begin{equation}
\mathcal{S}_r
=
\left\{
\boldsymbol{\theta}_{r}^{(q,f)}
\;:\;
q\in\mathcal{Q}_{\mathrm{basis}},\;
f=1,\ldots,F_q
\right\}
\label{eq:pooled_residue_samples}
\end{equation}
is computed in the $d_r$-dimensional local torsional conformation space \cite{rodriguez2018ComputingFreeEnergya,Glielmo2022DADApy}. The density peaks are then clustered using DPA \cite{derrico2021AutomaticTopographyHighdimensionala,Glielmo2022DADApy}, obtaining $K_r$ clusters. 
Note that the number of clusters $K_r$ is not fixed for each residue: it is identified automatically by the DPA algorithm based on the merging hyperparameter $Z$ (for which the value of $Z=3.0$ is used across all residues). These clusters define and label regions in the residue-specific torsional space via partition
\begin{equation}
\mathcal{C}_r = \{C_{r,1}, \, \ldots, \, C_{r,K_r}\}
\qquad \mathrm{s.t.} \qquad 
\bigcupdot_{k=1}^{K_r} C_{r,k} \,, 
\label{eq:residue_clusters}
\end{equation}
and are interpreted as local microstates. Therefore, for each frame $f$ of trajectory $q$, the residue $r$ is assigned a discrete state label $C_{r,k} \in \mathcal{C}_r$ that can be mapped to an integer-value coordinate
\begin{equation}
x_r^{(q,f)} = k \, \in \{1, \, \ldots, \, K_r\}\, 
\quad\mathrm{s.t.}\quad
\boldsymbol{\theta}_{r}^{(q,f)} \in C_{r,k}.
\label{eq:residue_state_label}
\end{equation}
The complete protein conformation therefore belongs to the heterogeneous product state space
\begin{equation}
\mathcal{X}
=
\prod_{r=1}^{N}\{1, \, \ldots, \, K_r\},
\label{eq:product_state_space}
\end{equation}
which has $N$ residue coordinates and cardinality $|\mathcal X|=\prod_{r=1}^{N}K_r$.
Each protein frame is represented by the discrete vector $\mathbf{x}^{(q,f)}=(x_1^{(q,f)},\ldots,x_N^{(q,f)}) \, \in \, \mathcal{X}$.
Importantly, the basis is shared across all trajectories: differences between functional states are therefore encoded by changes in residue-wise occupancies and inter-residue co-occurrence patterns, not by redefining the residue microstates themselves.

\begin{figure}[ht]
\captionsetup{font={small,stretch=1.0}}
\centering
\includegraphics[width=0.95\textwidth]{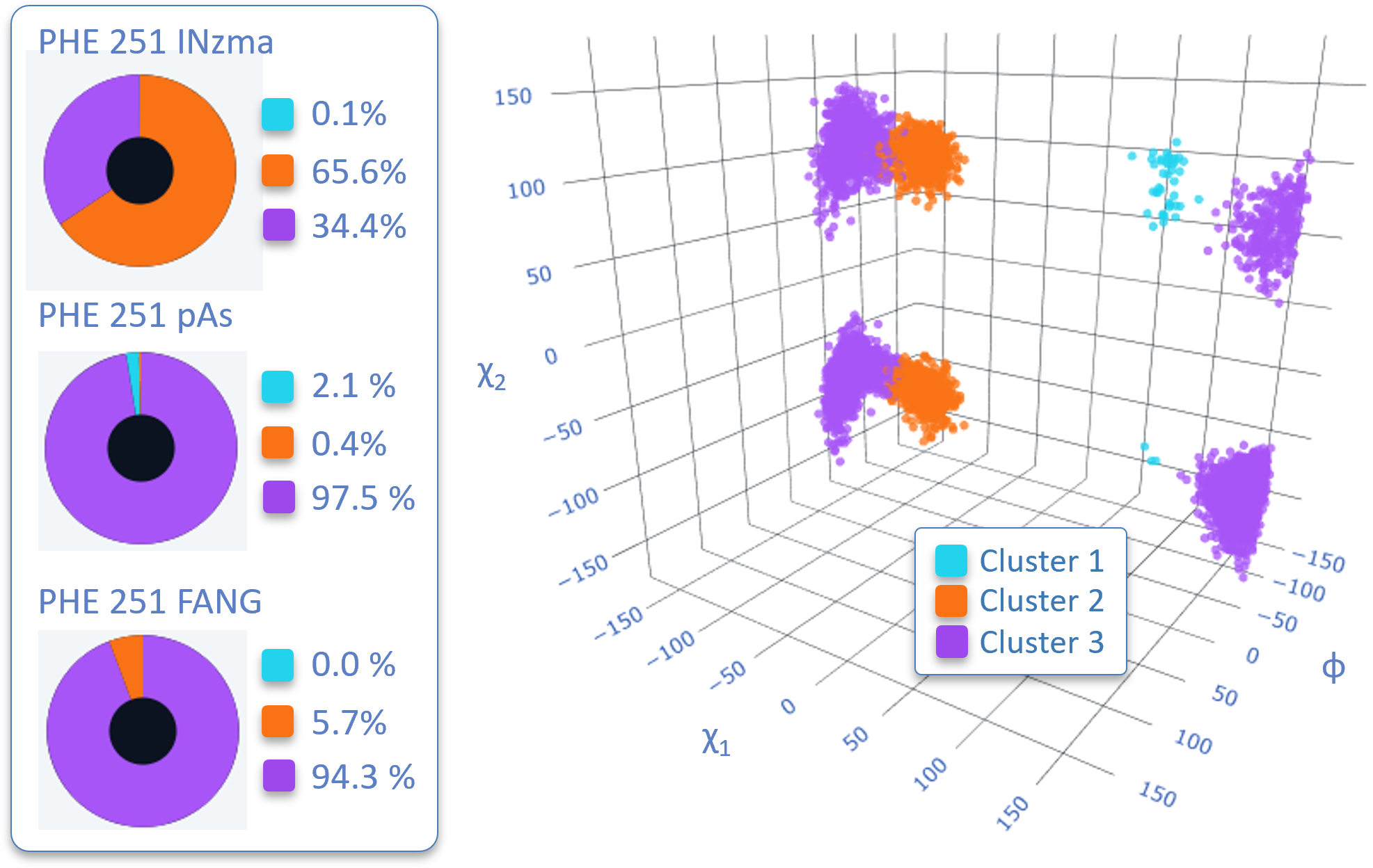}
\caption{\textbf{Example of a shared residue-wise microstate basis.}
A representative residue (here phenylalanine 251: PHE251) is shown in its local torsional feature space after periodic preprocessing of the relevant dihedral angles, including the symmetry-aware treatment of $\chi_2$.
The pooled frames from $\mathcal{Q}_{\mathrm{basis}}$ (INzma, pAs and FANG) are partitioned into discrete microstates (colours, right).
The pie charts (left) show the occupancies of the shared microstates in the inactive, pseudo-active (pAs), and active ensembles.}
\label{fig:residue_microstate_example}
\end{figure}

Figure~\ref{fig:residue_microstate_example} illustrates this construction for a representative residue.
All conformations of that residue observed in the trajectories belonging to $\mathcal{Q}_{\mathrm{basis}}$ are embedded in its local torsional space and clustered into a finite set of microstates. The same learned microstate basis is then used to describe different conformational ensembles. As shown by the population plots, the active, pseudo-active, and inactive ensembles differ in their occupancy frequencies of the shared microstates.

\subsection{Local residue-contact graph}

The pairwise terms of the Potts model are restricted to a sparse structural graph connecting residues that come into direct spatial proximity in the molecular-dynamics data.
Let \[ \mathcal{G}=(\mathcal{V},\mathcal{E}) \] denote this graph, where the vertex set $\mathcal{V}=\{1,\ldots,N\}$ contains the $N$ receptor residues.
For residue $r$, trajectory $q\in\mathcal{Q}_{\mathrm{basis}}$, and frame $f$, we denote the Cartesian position of its C$_\alpha$ atom by \[ \mathbf{c}_{r}^{(q,f)}\in\mathbb{R}^{3}. \]
Two residues $r$ and $s$ are connected whenever their C$_\alpha$ atoms approach within a prescribed cutoff $r_c$ in at least one frame of the basis trajectories.
Formally,

\begin{equation}
(r,s)\in\mathcal{E} \quad\Longleftrightarrow\quad \min_{\substack{ q\in\mathcal{Q}_{\mathrm{basis}}\\ f\in\{1,\ldots,F_q\} }} \left\| \mathbf{c}_{r}^{(q,f)} - \mathbf{c}_{s}^{(q,f)} \right\|_2 < r_c, \qquad r<s .
\label{eq:contact_graph}
\end{equation}

Equivalently, the graph contains the union of all residue--residue contacts observed across $\mathcal{Q}_{\mathrm{basis}}$.
A contact therefore needs to occur only transiently in one of the simulated ensembles to be retained in the graph.
This construction is deliberately more inclusive than defining contacts from a single reference structure: residues that interact only in one conformational regime can still be coupled in a common statistical model.
The graph topology is fixed before fitting the Potts Hamiltonian and is shared across the state-specific models considered in this work.
Consequently, changes between Hamiltonians arise from the learned statistical fields and couplings.
The cutoff $r_c$ controls the locality and complexity of the model.

\subsection{Sparse Potts Hamiltonian}
\label{subsec:potts_hamiltonian}

For a reduced protein configuration \(\mathbf{x}\in\mathcal{X}\), \PHASE{} defines the pairwise effective Hamiltonian
\begin{equation}
    E_{\theta}(\mathbf{x})
    =
    \sum_{r=1}^{N}h_r(x_r)
    +
    \sum_{(r,s)\in\mathcal{E}}J_{rs}(x_r,x_s),
    \label{eq:potts_energy_methods}
\end{equation}
where \(h_r\in\mathbb{R}^{K_r}\) is the field associated with residue \(r\), \(J_{rs}\in\mathbb{R}^{K_r\times K_s}\) is the coupling block associated with edge \((r,s)\), and
\begin{equation}
    \theta=\{h_r,J_{rs}\}_{r,(r,s)\in\mathcal{E}}
\end{equation}
denotes the complete parameter set. Each undirected edge is stored once with \(r<s\); when the opposite orientation is required, \(J_{sr}(\ell,k)=J_{rs}(k,\ell)\).

The Hamiltonian defines the reduced-space Boltzmann distribution
\begin{equation}
    p_{\theta,\beta}(\mathbf{x})
    =
    \frac{\exp[-\beta E_{\theta}(\mathbf{x})]}
    {Z_{\theta}(\beta)},
    \qquad
    Z_{\theta}(\beta)
    =
    \sum_{\mathbf{x}\in\mathcal{X}}
    \exp[-\beta E_{\theta}(\mathbf{x})].
    \label{eq:potts_distribution_methods}
\end{equation}
Fitting and target-distribution sampling use \(\beta=1\), unless specified otherwise. Accordingly, \(E_{\theta}\) is dimensionless and represents an effective negative-log statistical weight in the residue-microstate space.

\subsection{Pseudolikelihood fitting}

The fields and pairwise couplings of the Potts Hamiltonian are inferred from the discrete trajectory configurations by pseudolikelihood maximization (PLM) \cite{Besag1975,Ekeberg2013}. Direct maximum-likelihood estimation would require evaluating the partition function
\[
Z_\theta =
\sum_{\mathbf{x}\in\mathcal{X}}
\exp[-E_\theta(\mathbf{x})],
\]
whose cost grows exponentially with the number of residues. PLM avoids this global normalization by expressing the fitting objective in terms of the conditional probability of each residue given the current states of its neighbors.

For a configuration $\mathbf{x}=(x_1,\ldots,x_N)$ and residue $r$, the Potts Hamiltonian gives the conditional distribution
\begin{equation}
P_\theta
\left(
x_r=k \mid \mathbf{x}_{\setminus r}
\right)
=
\frac{
\exp\left[
-h_r(k)
-\displaystyle\sum_{s\in\partial r}
J_{rs}(k,x_s)
\right]
}{
\displaystyle\sum_{k'=1}^{K_r}
\exp\left[
-h_r(k')
-\displaystyle\sum_{s\in\partial r}
J_{rs}(k',x_s)
\right]
},
\label{eq:potts_conditional}
\end{equation}
where $\partial r$ denotes the residues connected to $r$ in the contact graph.
For an edge stored with the opposite orientation, the corresponding coupling matrix is transposed so that its first index always refers to the state of residue $r$.

Given a training set of $T$ discrete configurations $\{\mathbf{x}^{(t)}\}_{t=1}^{T}$, the log-pseudolikelihood is
\begin{equation}
\mathcal{L}_{\mathrm{PL}}
(\theta)
=
\sum_{t=1}^{T}
\sum_{r=1}^{N}
\log
P_\theta
\left(
x_r^{(t)}
\mid
\mathbf{x}_{\setminus r}^{(t)}
\right).
\label{eq:potts_pl}
\end{equation}

The optimization is initialized from empirical single-residue and pairwise
statistics. For residue $r$ and edge $(r,s)$, let $p_r(k)$ and
$p_{rs}(k,l)$ denote the empirical marginal and joint probabilities. The
initial parameters are constructed as
\begin{align}
h_r^{(0)}(k)
&=
-\log p_r(k),\\
J_{rs}^{(0)}(k,l)
&=
-\log
\frac{
p_{rs}(k,l)
}{
p_r(k)p_s(l)
},
\end{align}
with a small probability regularization to avoid logarithms of zero.
These pointwise-mutual-information estimates provide an initialization only; the final Potts Hamiltonian is obtained by optimizing the regularized pseudolikelihood objective
\begin{equation}
\mathcal{L}_{\mathrm{PLM}}
=
-\sum_{\mathbf{x}\in\mathcal{D}}
\sum_{r=1}^{N}
\log
P_\theta
\left(
x_r
\mid
\mathbf{x}_{\setminus r}
\right)
+
\lambda_2
\left[
\sum_r \|h_r\|_2^2
+
\sum_{(r,s)\in\mathcal{E}}
\|J_{rs}\|_F^2
\right]
+
\lambda_J
\sum_{(r,s)\in\mathcal{E}}
\|J_{rs}\|_F .
\label{eq:plm_objective}
\end{equation}
The $L_2$ term controls the overall magnitude of the fields and couplings, whereas the group-Frobenius penalty acts on complete coupling blocks and preferentially suppresses weak residue--residue interactions.

The objective is evaluated over minibatches of trajectory frames and optimized using AdamW with a cosine learning-rate schedule for 1000 epochs and a batch size of 8192. The learning rate is decreased from $10^{-2}$ to $10^{-3}$, with $\lambda_2=10^{-5}$ and $\lambda_J=10^{-3}$. Parameters are represented in the zero-sum gauge, which is applied consistently during optimization, regularization and model export.

\subsection{Zero-sum gauge normalization}
\label{subsec:zero_sum_gauge}

The numerical values of the Potts fields and couplings are not uniquely defined.
Part of the contribution associated with a pairwise coupling can be shifted into the corresponding single-residue fields without changing the probability distribution represented by the model.
Consequently, two equivalent Potts parameterizations can assign different numerical values to $h_r$ and $J_{rs}$ even though they describe the same ensemble.

To obtain a consistent and interpretable parameterization, all fitted Hamiltonians are expressed in the zero-sum gauge.
For an interacting residue pair $(r,s)$, we define the mean of each row, the mean of each column, and the overall mean of the coupling matrix as

\begin{align}
\overline{J}_{rs}^{(r)}(k)
&=
\frac{1}{K_s}
\sum_{\ell=1}^{K_s}
J_{rs}(k,\ell),
\\
\overline{J}_{rs}^{(s)}(\ell)
&=
\frac{1}{K_r}
\sum_{k=1}^{K_r}
J_{rs}(k,\ell),
\\
\overline{J}_{rs}
&=
\frac{1}{K_rK_s}
\sum_{k=1}^{K_r}
\sum_{\ell=1}^{K_s}
J_{rs}(k,\ell).
\end{align}

The gauge-normalized coupling is obtained by subtracting its row and column means and restoring the overall mean,

\begin{equation}
\widetilde{J}_{rs}(k,\ell)
=
J_{rs}(k,\ell)
-
\overline{J}_{rs}^{(r)}(k)
-
\overline{J}_{rs}^{(s)}(\ell)
+
\overline{J}_{rs}.
\label{eq:gauge_coupling_methods}
\end{equation}

The contributions removed from the coupling matrices are transferred to the single-residue fields.
For residue $r$, we first define
\begin{equation}
h_r^{*}(k)
=
h_r(k)
+
\sum_{s\in\partial r}
\overline{J}_{rs}^{(r)}(k),
\label{eq:gauge_field_intermediate}
\end{equation}
where $\partial r$ denotes the neighbours of residue $r$ in the contact graph and $J_{rs}$ is written with its first index corresponding to residue $r$.
The field is then centered over its $K_r$ microstates,

\begin{equation}
\widetilde{h}_r(k)
=
h_r^{*}(k)
-
\frac{1}{K_r}
\sum_{k'=1}^{K_r} h_r^{*}(k').
\label{eq:gauge_field_methods}
\end{equation}

The resulting parameters satisfy

\begin{equation}
\sum_{k=1}^{K_r}\widetilde{h}_r(k)=0,
\qquad
\sum_{k=1}^{K_r}\widetilde{J}_{rs}(k,\ell)=0,
\qquad
\sum_{\ell=1}^{K_s}\widetilde{J}_{rs}(k,\ell)=0.
\label{eq:gauge_conditions_methods}
\end{equation}

Intuitively, the zero-sum gauge assigns the average contribution of each pair interaction to the corresponding single-residue fields, leaving ${J}_{rs}$ to describe only state-specific deviations from those averages.
This transformation changes the absolute energy only by a configuration-independent constant and therefore leaves all probabilities and energy differences within the fitted model unchanged.
Its purpose is to place independently fitted Potts models in the same parameter convention before their fields, couplings, or energies are compared.

\subsection{Endpoint preference score}
\label{subsec:endpoint_preference}

To compare individual conformations with the two stabilized endpoints of the \AtwoA{} activation landscape, separate Potts Hamiltonians are fitted to the INzma and FANG ensembles, representing the full inactive and full active reference states, respectively.
Both models use the same residue microstate basis and contact graph and are expressed in the zero-sum gauge described above.

For a discrete protein configuration $\mathbf{x}$, we define the endpoint preference score as reported in Eq. \ref{eq:delta_endpoint_results}.
Negative values indicate that the configuration receives a lower energy under the INzma Hamiltonian, whereas positive values indicate a lower energy under the FANG Hamiltonian.
The score can therefore be used to order configurations according to their relative compatibility with the two learned endpoint ensembles.

To spatially resolve this preference, the same score is evaluated on selected residue subsets while retaining their interactions with the surrounding protein.
For a residue set $R$, we define the corresponding regional energy under model $M$ as

\begin{equation}
\widetilde{E}_{M}^{(R)}(\mathbf{x}) = \sum_{i\in R} \widetilde{h}_{i}^{M}(x_i) + \sum_{\substack{(i,j)\in\mathcal{E}\\ i\in R\,\lor\,j\in R}} \widetilde{J}_{ij}^{M}(x_i,x_j),
\label{eq:regional_energy_methods}
\end{equation}

where all pairwise terms incident on at least one residue in $R$ are retained, including couplings to residues outside the selected region.
The regional endpoint preference is then

\begin{equation}
\Delta E_{\mathrm{INzma-FANG}}^{(R)}(\mathbf{x}) = \widetilde{E}_{\mathrm{INzma}}^{(R)}(\mathbf{x}) - \widetilde{E}_{\mathrm{FANG}}^{(R)}(\mathbf{x}).
\label{eq:regional_endpoint_score_methods}
\end{equation}

This definition preserves the complete interaction profile of the selected residues.
For the spatial decomposition of A$_{2A}$, we evaluate Eq.~\eqref{eq:regional_endpoint_score_methods} separately for the extracellular region, comprising the orthosteric pocket and extracellular loops, and for the intracellular region, comprising the cytoplasmic portions of the transmembrane helices and intracellular loops.
The resulting $\Delta E_{\mathrm{extra}}$ and $\Delta E_{\mathrm{intra}}$ values are analyzed both independently and jointly to compare the ligand-facing and transducer-facing conformational signatures of each ensemble.

Importantly, $\Delta E_{\mathrm{INzma-FANG}}$ and its regional counterparts are relative statistical preference scores and not physical free-energy difference between the inactive and active macrostates.
For each model $M$,

\begin{equation}
p_M(\mathbf{x})=\frac{\exp[-E_M(\mathbf{x})]}{Z_M},
\end{equation}

and therefore

\begin{equation}
\log
\frac{p_{\mathrm{FANG}}(\mathbf{x})}
     {p_{\mathrm{INzma}}(\mathbf{x})}
=
\Delta E_{\mathrm{INzma-FANG}}(\mathbf{x})
+
\log
\frac{Z_{\mathrm{INzma}}}{Z_{\mathrm{FANG}}}.
\end{equation}

The partition functions are not evaluated in this work.
Their contribution, however, is independent of $\mathbf{x}$ and therefore only shifts the endpoint score by a constant.
Consequently, $\Delta E_{\mathrm{INzma-FANG}}$ preserves the relative ordering of configurations according to their preference for the two endpoint models, although its absolute zero cannot be interpreted as the point of equal endpoint probability.

\subsection{Gibbs and replica-exchange sampling}
\label{subsec:replica_exchange_sampling}

Once the Potts Hamiltonian has been fitted, configurations must be sampled from the corresponding probability distribution in order to generate new protein microstate combinations and to assess whether the learned model reproduces the statistics of the molecular-dynamics ensemble. Gibbs sampling provides a natural sampling procedure for this purpose because the conditional distribution of each residue is available analytically from the Potts Hamiltonian~\cite{CasellaGeorge1992}. Consequently, sampling does not require evaluation of the global partition function.

At inverse temperature $\beta$, the Potts model defines
\begin{equation}
    p_{\theta,\beta}(\mathbf{x})
    =
    \frac{
    \exp[-\beta E_{\theta}(\mathbf{x})]
    }{
    Z_{\theta}(\beta)
    }.
    \label{eq:tempered_potts_distribution}
\end{equation}
The fitted ensemble corresponds to $\beta=1$. Values $\beta<1$ are used only for tempering during replica-exchange sampling and should not be interpreted as physical temperatures of the molecular system.

Starting from a configuration $\mathbf{x}$, a Gibbs update selects one residue $r$ while keeping all other residue states fixed. Its new state is drawn directly from the conditional distribution $p_{\theta,\beta}(x_r\mid\mathbf{x}_{\setminus r})$ defined in Eq.~\eqref{eq:potts_conditional}. Because the Potts Hamiltonian is pairwise and sparse, this conditional probability depends only on the current states of the neighbours of $r$ in the contact graph. One Gibbs sweep consists of updating every residue once.

Under the usual ergodicity conditions, repeated Gibbs updates converge to $p_{\theta,\beta}(\mathbf{x})$, which is therefore the stationary distribution of the Markov chain. At $\beta=1$, sufficiently equilibrated Gibbs samples are thus samples from the fitted Potts model. Importantly, this does not by itself imply agreement with the original MD ensemble: comparison of the sampled and
MD distributions provides a direct validation of how accurately the fitted pairwise Hamiltonian represents the trajectory data.

For protein ensembles, however, the learned distribution can contain well-separated high-probability regions. Moving between them through single-residue updates may require visiting low-probability intermediate configurations, resulting in slow mixing of a single Gibbs chain. We therefore use replica-exchange sampling to facilitate exploration of the discrete configuration space~\cite{HukushimaNemoto1996}. Multiple Gibbs chains are run in parallel at inverse temperatures
\begin{equation}
    0 < \beta_1 < \beta_2 < \cdots < \beta_L = 1,
    \label{eq:replica_beta_ladder}
\end{equation}
Lower values of $\beta$ flatten the energy landscape and allow the corresponding replicas to move more readily between otherwise weakly connected regions of configuration space.

In all replica-exchange calculations, we use $L=8$ replicas with inverse temperatures geometrically spaced between $\beta_{\min}=0.2$ and $\beta_{\max}=1.0$, including both endpoints. The ladder is therefore defined as

\begin{equation}
    \beta_m
    =
    \beta_{\min}
    \left(
    \frac{\beta_{\max}}{\beta_{\min}}
    \right)^{\frac{m-1}{L-1}},
    \qquad
    m=1,\ldots,L.
    \label{eq:geometric_beta_ladder}
\end{equation}

After each replica-exchange round, neighbouring replicas may exchange their configurations. For adjacent replicas $m$ and $m+1$, containing configurations $\mathbf{x}_m$ and $\mathbf{x}_{m+1}$, the exchange is accepted with probability

\begin{equation}
    \alpha_{m,m+1}
    =
    \min\left\{
    1,
    \exp\left[
    (\beta_m-\beta_{m+1})
    \left(
    E_{\theta}(\mathbf{x}_m)
    -
    E_{\theta}(\mathbf{x}_{m+1})
    \right)
    \right]
    \right\}.
    \label{eq:replica_exchange_methods}
\end{equation}

This exchange criterion preserves the target distribution associated with each replica while allowing configurations to diffuse across the temperature ladder.

After discarding the first 100 replica-exchange rounds as burn-in, samples are retained every two rounds.
This yields $2,000$ retained configurations per replica for each sampling run.
Only samples from the target replica at $\beta=1$ are used for subsequent analyses.

\subsection{QUBO encoding and classical simulated annealing}
\label{subsec:qubo}

The discrete Potts Hamiltonian can also be expressed as a \emph{quadratic unconstrained binary optimization} (QUBO) problem, a standard quadratic representation over binary variables that can be explored using classical or quantum annealing algorithms.
To obtain this representation, we encode the discrete microstate of each residue using one-hot binary variables.

For each residue $r$ and microstate $k\in\{1,\ldots,K_r\}$, we introduce a
binary variable
\begin{equation}
    y_{r,k}\in\{0,1\},
\end{equation}
where $y_{r,k}=1$ indicates that residue $r$ occupies microstate $k$. A valid
Potts configuration therefore satisfies the one-hot constraint
\begin{equation}
    \sum_{k=1}^{K_r} y_{r,k}=1
    \qquad \forall r.
    \label{eq:qubo_onehot}
\end{equation}

The Potts energy can then be written as
\begin{equation}
E_{\mathrm{Potts}}(\mathbf y)
=
\sum_{r=1}^{N}\sum_{k=1}^{K_r}
h_r(k)y_{r,k}
+
\sum_{(r,s)\in\mathcal E}
\sum_{k=1}^{K_r}\sum_{\ell=1}^{K_s}
J_{rs}(k,\ell)y_{r,k}y_{s,\ell}.
\label{eq:binary_potts_energy}
\end{equation}

The one-hot constraints are incorporated into the binary objective through quadratic penalty terms,

\begin{equation}
E_{\mathrm{QUBO}}(\mathbf y)
=
\beta E_{\mathrm{Potts}}(\mathbf y)
+
\sum_{r=1}^{N}
\lambda_r
\left(
\sum_{k=1}^{K_r}y_{r,k}-1
\right)^2.
\label{eq:qubo_methods}
\end{equation}

For every valid one-hot configuration the penalty term vanishes, so that the QUBO preserves the Potts energy landscape, up to the overall factor $\beta$, within the feasible subspace.

To account for the heterogeneous magnitude of the local Potts parameters, we define a residue-dependent reference penalty
\begin{equation}
\lambda_r^{(0)}
=
\max\left[
10^{-6},
\max_k |h_r(k)|
+
\sum_{s\in\partial r}
\max_{k,\ell}|J_{rs}(k,\ell)|
\right],
\label{eq:qubo_base_penalty}
\end{equation}
and scale it globally as
\begin{equation}
\lambda_r=s\,\lambda_r^{(0)}.
\label{eq:qubo_penalty_scale}
\end{equation}
The scalar $s$ therefore controls the relative strength of the one-hot constraint without changing the relative penalty scale among residues.

For the classical annealing baseline, we evaluate the INapo 6~\AA{} Hamiltonian under four global penalty scales,

\begin{equation}
s \in \{1,\;1.5,\;2,\;4\},
\end{equation}

where $\lambda_r = s\,\lambda_r^{(0)}$ and $\lambda_r^{(0)}$ is defined in Eq.~\eqref{eq:qubo_base_penalty}.
For each penalty scale, we perform 2000 independent simulated-annealing reads initialized from valid microstate configurations drawn from the INapo MD ensemble (\emph{warm start}).
The annealing schedule is held fixed to a geometric progression from $\beta=0.1$ to $\beta=2.0$ across the penalty sweep. We run 5000 sweeps per read.

Each returned binary configuration is then checked for one-hot validity.
Configurations violating the one-hot constraint for one or more residues are classified as invalid and excluded from the Potts-energy distributions.
Among valid configurations, we evaluate the decoded Potts energy $E_{\mathrm{Potts}}(\mathbf{x})$ and compare its distribution with the reference MD ensemble and with replica-exchange Gibbs samples drawn from the same fitted Hamiltonian.

To assess whether annealing remains trapped near the supplied MD warm start or explores distinct valid configurations, we additionally compute, for each valid read, the Hamming distance between the decoded final microstate vector and the exact microstate vector used as initialization.

For a valid annealing read initialized from $\mathbf{x}^{(0)}$ and ending at $\mathbf{x}^{(\mathrm{SA})}$, the Hamming distance is

\begin{equation}
d_H\!\left(\mathbf{x}^{(\mathrm{SA})},\mathbf{x}^{(0)}\right)=\sum_{r=1}^{N}\mathbb{I}\!\left[x_r^{(\mathrm{SA})}\neq x_r^{(0)}\right].
\end{equation}

The corresponding changed fraction was defined as

\begin{equation}
f_{\mathrm{chg}}
=
\frac{1}{N}
d_H\!\left(\mathbf{x}^{(\mathrm{SA})},\mathbf{x}^{(0)}\right).
\end{equation}

\begin{figure}[ht]
    \captionsetup{font={small,stretch=1.0}}
    \centering
    \includegraphics[width=\linewidth]{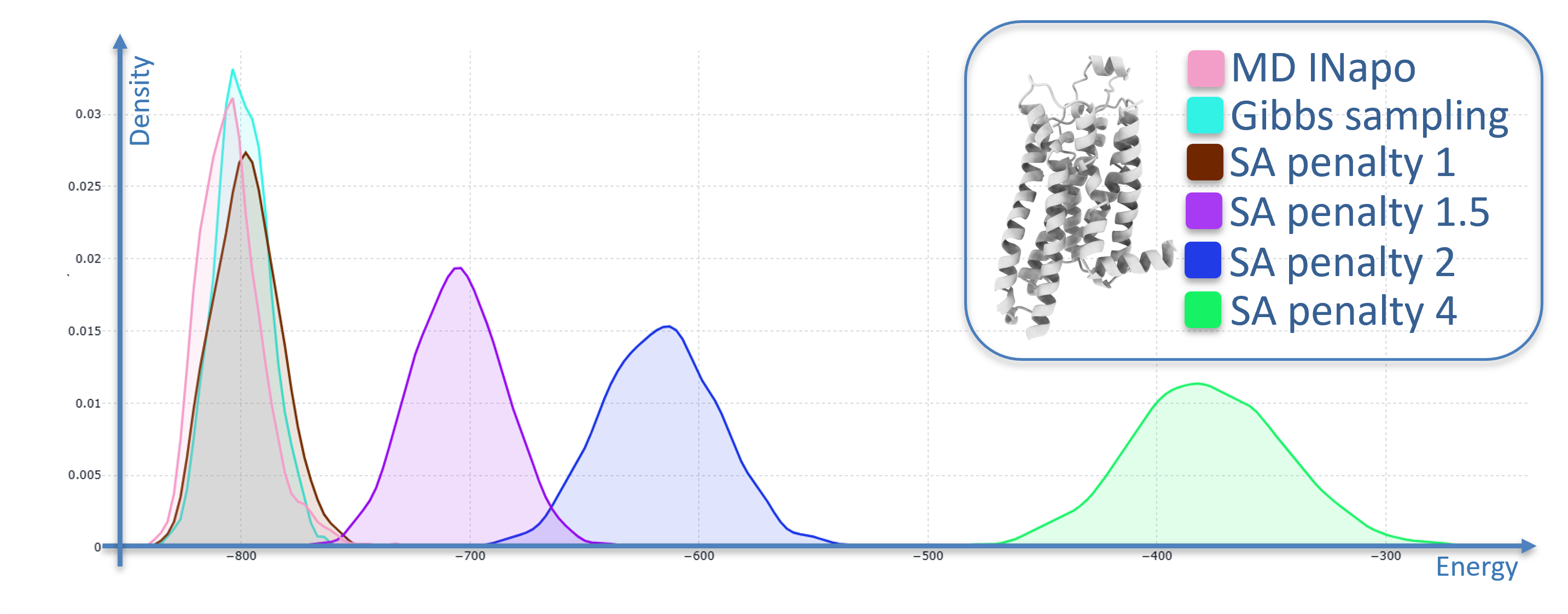}
    \caption{\textbf{Classical simulated annealing baseline for QUBO exploration of the INapo \PHASE{} Hamiltonian.}
    Potts-energy distributions are shown for the reference INapo MD ensemble, replica-exchange Gibbs samples from the fitted 6~\AA{} Hamiltonian, and valid configurations obtained by simulated annealing of its one-hot QUBO
    representation using penalty scales $s=1$, $1.5$, $2$, and $4$.
    Only configurations satisfying the one-hot constraints for all residues are included in the simulated-annealing distributions. Simulated annealing is initialized from configurations drawn from the INapo MD ensemble.
    The resulting penalty-dependent validity and movement away from the initialization are reported in Table~\ref{tab:qubo_sa_diagnostics}.}
    \label{fig:qubo_sa_baseline}
\end{figure}

The resulting Potts-energy distributions and sampling diagnostics are summarized in Fig.~\ref{fig:qubo_sa_baseline} and Table~\ref{tab:qubo_sa_diagnostics}, respectively.

\begin{table}[t]
\centering
\small
\caption{\textbf{Diagnostics of Gibbs and simulated-annealing sampling.}
The valid fraction is the fraction of returned configurations satisfying all one-hot constraints.
The changed-residue fraction is computed relative to the MD configuration used to initialize each simulated-annealing read and is evaluated over valid outputs only.
Replica-exchange Gibbs sampling operates directly in the valid Potts state space; no warm-start movement diagnostic is therefore defined.}
\label{tab:qubo_sa_diagnostics}

\begin{tabular}{lcc}
\toprule
Sampling method & Valid fraction & Changed-residue fraction \\
\midrule
Gibbs sampling & 1.000 & -- \\
SA, $s=1$   & 789/2000  & 0.291 \\
SA, $s=1.5$ & 1349/2000 & 0.348 \\
SA, $s=2$   & 1525/2000 & 0.396 \\
SA, $s=4$   & 1535/2000 & 0.509 \\
\bottomrule
\end{tabular}
\end{table}

\subsection{Computational implementation}

All calculations were performed on a NVIDIA RTX A6000 GPU.
Fitting a single PHASE Hamiltonian required approximately $3$ hours, replica-exchange Gibbs sampling of $2,000$ configurations required approximately $30$ seconds using parallel workers, one per replica. Simulated annealing sampling requires even less time, finishing in just few seconds.

\subsection{Marginal-distribution validation}
\label{subsec:marginal_validation}

To quantify how accurately samples from a fitted Potts Hamiltonian reproduce the MD ensemble, we compare their residue-wise and residue-pair state distributions using the Jensen--Shannon (JS) divergence.

For each residue $r$, we estimate the empirical microstate distribution $p_r^{\mathrm{MD}}(k)$ from the MD trajectory and the corresponding distribution $p_r^{\mathrm{Potts}}(k)$ from configurations sampled from the fitted model.
This comparison tests whether the model reproduces the population of the individual microstates defined by the shared residue-wise basis.

We then perform the analogous comparison for pairs of residues. For every residue pair $(r,s)$, irrespective of whether $(r,s)$ belongs to the Potts contact graph, we estimate the joint microstate distribution $p_{rs}(k,\ell)$ and compare the MD and sampled distributions.
Importantly, the Hamiltonian contains an explicit coupling $J_{rs}$ only when $(r,s)\in\mathcal{E}$, whereas the validation is performed over all residue pairs. Pairwise agreement outside $\mathcal{E}$ therefore tests whether correlations not represented by a direct coupling are nevertheless recovered through the network of local interactions.

For two categorical distributions $p$ and $q$, the Jensen--Shannon divergence is

\begin{equation}
    \operatorname{JS}(p\Vert q)
    =
    \frac{1}{2}\operatorname{KL}(p\Vert m)
    +
    \frac{1}{2}\operatorname{KL}(q\Vert m),
    \qquad
    m=\frac{p+q}{2},
    \label{eq:js_methods}
\end{equation}

with

\begin{equation}
    \operatorname{KL}(p\Vert m)
    =
    \sum_i p_i\log\frac{p_i}{m_i}.
\end{equation}

Natural logarithms are used. Before evaluation, probabilities are clipped to $10^{-12}$ and renormalized to avoid numerical singularities for unobserved states. A JS divergence of zero indicates identical distributions, while increasing values indicate progressively larger differences between the MD and Potts-sampled ensembles.

For the cutoff analysis shown in the main text, this validation is performed on the INapo ensemble for Potts contact graphs constructed with $C_\alpha$ cutoffs of 4, 5, 6, and 10~\AA{}. Residue-wise JS divergences are reported individually for all residues, while residue-pair JS divergences are shown as complete $N\times N$ matrices. The same residue-wise validation at the selected 6~\AA{} cutoff is reported for the remaining \AtwoA{} ensembles in the Supplementary Information.

\subsection{Cluster-conditioned all-atom backmapping}
\label{subsec:cluster_conditioned_backmapping}
With the backmapping model we perform atomistic reconstruction from microstate cluster configuration $\mathbf{x}$ to protein coordinates $\mathbf{R}$.
Since cluster labels are residue-specific, the pair $(r,x_r)$ is mapped to a unique global token. We define the offset $o_r$ of residue $r$ as

\begin{equation}
    o_r
    =
    \sum_{j=1}^{r-1} K_j,
    \qquad
    g_r
    =
    o_r+x_r,
    \qquad
    g_r\in
    \left\{1,\ldots,\sum_{j=1}^{N}K_j\right\}.
    \label{eq:global_cluster_id_methods}
\end{equation}

\noindent We fine-tune a protein-folding model using the unique global token vector $\mathbf{g}=(g_1,\ldots,g_N)$ as conditioning.
We select the SimpleFold model \cite{Wang2025SimpleFold}, a general-purpose flow-matching protein-folding model built solely from transformer blocks with adaptive layers. It does not present any domain-specific inductive biases and components—such as triangular updates, explicit pair representations, curated training objectives, MSAs, contact maps, and equivariant geometric modules—to reduce architectural complexity.
The checkpoint from which we start the fine-tuning is a SimpleFold model with 100M parameters that has been trained end-to-end on approximately 9M distilled protein structures together with experimental PDB data.
We fine-tune on 16,000 structures from a subsampling  of \AtwoA{} trajectories (INzma, INapo, Theo and FANG, N=4,000 per trajectory) using disjoint training and test splits. The reported test set contained 1,000 structures per  conformational ensemble. 
The SimpleFold model $\mathcal{B}_{\zeta}$ takes the all protein heavy-atoms coordinates, the amino acid sequence $\mathbf{s}$, the timestep embedding \textit{t} and the global token vector $\mathbf{g}$ to model the target velocity $v_t = \mathbf{R} - \epsilon$, where $\mathbf{R}\in\mathbb{R}^{A\times3}$ is an all-heavy-atom receptor structure and $\epsilon \sim N(0, I) \in \mathbb{R}^{A\times3}$. The flow-matching objective is defined as follows \cite{albergo2025stochasticinterpolantsunifyingframework,albergo2023buildingnormalizingflowsstochastic}
$$\mathcal{L}_{FM} = \mathbf{E}_{\mathbf{R},s,\epsilon,t,\textbf{g}}  \left[\frac{1}{|\mathcal{A}|}\|\mathcal{B}_{\zeta}(\mathbf{R}_t, s, t, \textbf{g}) - (\mathbf{R} - \epsilon)\|^{2} \right]$$
\noindent where $\mathbf{R}_t$ is a “noisy“ structure sampled from a linear interpolant path $\mathbf{R}_t = t\mathbf{R} + (1 - t) \epsilon$. 
To encourage the model to learn the influence of the conditioning signal $\mathbf{g}$, we mask some (15\%) of its entries and train a simple projection head to reconstruct them from a hidden residue-wise representation using the CrossEntropy loss function as in the Masked Language Model \cite{devlin2019bertpretrainingdeepbidirectional}
$$\mathcal{L}_{MLM}=-\sum_{g_{i} \in mask} log(p_{\zeta}(g_i | \tilde{\textbf{g}}))$$
where $g_i$ is the original cluster token and $\mathbf{\tilde{g}}$ is the corrupted global token, thus obtaining a final loss function as 
$$\mathcal{L}=\mathcal{L}_{FM} + \alpha\mathcal{L}_{MLM}$$
We set $\alpha = 0.05$.
Furthermore, we employ the Classifier-Free Guidance (CFG) technique \cite{ho2022classifierfreediffusionguidance} to sample with conditioning. The velocity field learnt is thus
$$ v_\zeta(\mathbf{R},s,t,\textbf{g}) =  v_\zeta(\mathbf{R},s,t,\textbf{g}=\emptyset) + \gamma(v_\zeta(\mathbf{R},s,t,\textbf{g})-v_\zeta(\mathbf{R},s,t,\textbf{g}=\emptyset))$$
from which we can compute
$$s_\zeta(\mathbf{R},s,t,\textbf{g}) = (t v_\zeta(\mathbf{R},s,t,\textbf{g}) - \mathbf{R})/ (1-t)$$
considering that we start from pretrained weights the conditioning is dropped out (i.e. set to $\textbf{g}=\emptyset$) with probability 0.1.

We perform sampling by initializing $\mathbf{R}_{0} \sim N (0, I)$ and integrate the learned vector field from $t=0$ to $t=1$ using a Langevin-style SDE formulation of the flow process (as done in SimpleFold), but we modify the update rule in the Euler–Maruyama integrator \cite{ma2024sitexploringflowdiffusionbased} such to consider the CFG contribution
$$ d\mathbf{R}_t = v_\zeta(\mathbf{R},s,t,\textbf{g}) dt + \frac{1}{2} w(t)s_\zeta(\mathbf{R},s,t,\textbf{g})dt+\sqrt{\tau w(t)}d \mathbf{\bar{W}}_t $$
where $w(t) > 0$ is a time-dependent diffusion coefficient, $\mathbf{\bar{W}}_t$ is a reverse-time Wiener process, and $\tau$ controls the scale of stochasticity.
To select $\tau$ and the guidance scale $\gamma$ we perform a grid-search as presented in \S \ref{sec:backmapping_res}.

\begin{figure}[ht]
\captionsetup{font={small,stretch=1.0}}
\centering
\includegraphics[width=\textwidth]{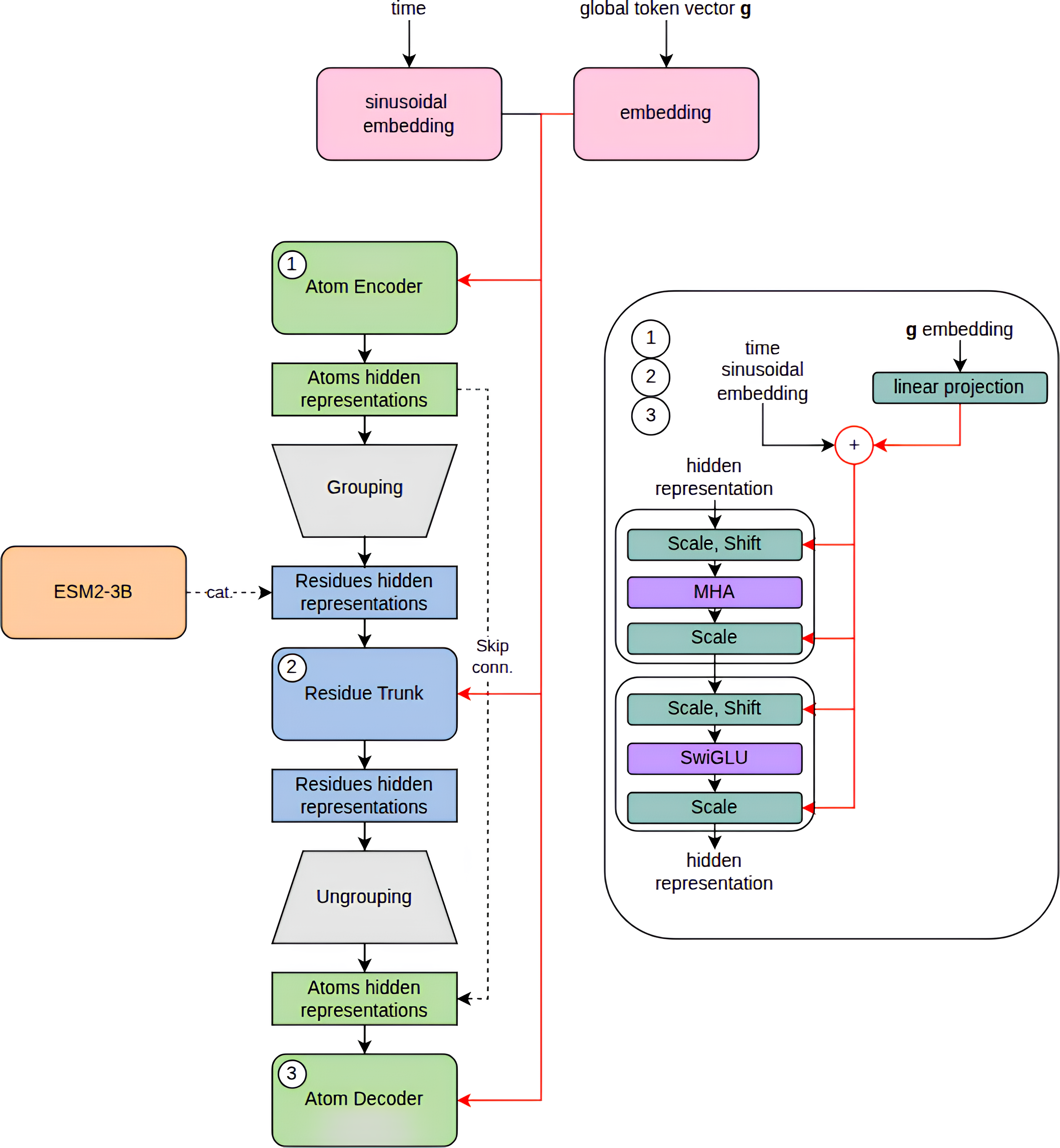}
\caption{\textbf{Backmapping model architecture, in red our modifications to the SimpleFold method.} The grouping operation is a residue-wise average pooling of the atom representations, while the ungrouping is the broadcasting of residue representations. SO(3) equivariance is learnt via data augmentation.}
\label{fig:backmapping_architecture}
\end{figure}

From an architectural perspective SimpleFold is composed by 3 major modules atom encoder and decoder and a core residue trunk. All modules are implemented with Diffusion Transformers (DiT) \cite{peebles2023scalablediffusionmodelstransformers} with Adaptive Layer Normalization (AdaLN) \cite{peebles2023scalablediffusionmodelstransformers} where the modulation of the input tokens is conditioned not only the timestep \textit{t} but also on a processed version of the embedding global token vector \textbf{g}. In Fig. \ref{fig:backmapping_architecture} it is possible to see highlighted in red our modifications to the original SimpleFold architecture.
We use an effective batch size of 16, where for each batch we select a single structure and we sample 16 different flow timestep \textit{t}. We fine-tune using the AdamW optimizer \cite{loshchilov2019decoupledweightdecayregularization} with a LinearWarmupCosineAnnealing for the learning rate with a minimum of $5\times10^{-7}$ and a maximum of $1\times10^{-4}$ to be reached after 10,000 steps. We train for 900,000 steps (56 epochs) and we reach the minimum learning rate after 850,000 steps.
We do not use cropping and the model dimensions are defined by the SimpleFold checkpoint selected. Also the input pipeline to featurize the initial atom representations does not deviate from the original method. We do not fine-tune the pLDDT head in our pipeline. Finally, as post-processing procedure, since the SimpleFold model is not equivariant w.r.t. reflections we have to perform a chirality sanity check to ensure that all residues have correct stereochemistry, and, if needed, fix the backmapped structure via reflection. We fine-tune on a single NVIDIA RTX A6000 for 23 days and the sampling of a single structure takes around 10 seconds.

\subsection{Backmapping evaluation and cycle consistency}
\label{subsec:backmapping_evaluation}

\noindent For every structure in the test set, we provide to the backmapping model its conditioning vector \textbf{g*}, and we use it to retrieve an associated reconstructed structure $\hat{\textbf{R}}$ as illustrated in \S \ref{subsec:cluster_conditioned_backmapping}. We then compute for each residue $r$ the microstate clusters of the reconstructed structure $\widehat{x}_r$ via density interpolation \cite{carli2021StatisticallyUnbiasedFreea} and cluster prediction \cite{ding2025CoupledReactionDiffusiona}, and compute the corresponding global token vector $\hat{\textbf{g}}$. 
We can therefore evaluate the quality of the backmapping model by computing
\begin{itemize}
    \item the RMSD between the original structure $\textbf{R*}$ and the backmapped structure $\hat{\textbf{R}}$,
    \item the percentage of microstate mismatch between $\textbf{g}^{*}$ and $\hat{\textbf{g}}$ 
\end{itemize}
For a atom subset $\mathcal{A}\subseteq\{1,\ldots,A\}$, the RMSD is computed via the Kabsch algorithm \cite{Kabsch1976} for alignment as
\begin{equation}
    \operatorname{RMSD}_{\mathcal{A}}
    =
    \min_{\mathbf{Q}\in SO(3),\,\mathbf{t}\in\mathbb{R}^3}
    \sqrt{
    \frac{1}{|\mathcal{A}|}
    \sum_{i\in\mathcal{A}}
    \left\|
    \mathbf{Q}\widehat{\mathbf{r}}_i
    +\mathbf{t}
    -\mathbf{r}_{i}^{*}
    \right\|_2^2
    }.
    \label{eq:rmsd_methods}
\end{equation}
The RMSD is computed for C$\alpha$, backbone, side-chain heavy-atoms and all-heavy-atoms. 
For both the RMSD calculation and the percentage of microstate mismatch, we also use an \AtwoA{} ad-hoc selection criteria to evaluate the described metrics only on residues of the structured region of the receptor, to avoid spurious contribution in our analysis due to the random coils and unstructured regions.

Given a residue subset $\mathcal{S}\subseteq\mathcal{V}$, the microstate mismatch count is defined as
\begin{equation}
    M_{\mathcal{S}}
    =
    \sum_{r\in\mathcal{S}}
    \mathbb{I}[\widehat{x}_r\neq x_{r}^{*}].
    \label{eq:cluster_mismatch_methods}
\end{equation}
The subset $\mathcal{S}$ is either the complete modeled receptor or the structured regions residue selection. We report in \S \ref{sec:backmapping_res} the percentage of microstate mismatch over the total residues selected. For newly sampled Potts configurations, no unique atomistic target exists; these structures were therefore evaluated by cycle consistency through the evaluation of microstate mismatch (i.e.  Eq.~\eqref{eq:cluster_mismatch_methods}).


\section*{Data availability}
The MD simulation dataset used in this study was originally generated by D'Amore et al. \cite{DAmore2024Chem} and is openly available on Zenodo at \url{10.5281/zenodo.13460724}.

\section*{Code availability}
The code developed for this study is openly available on GitHub at \url{https://github.com/limresgrp/PHASE}.

\section*{Acknowledgements}
The authors thank Stefano Raniolo and Samuele Di Cristofano for insightful discussions and suggestions during the preparation of this manuscript.


\section*{Competing interests}
The authors declare no competing interests.

\printbibliography

\clearpage


\setcounter{section}{0}
\renewcommand{\thesection}{S\arabic{section}}

\setcounter{figure}{0}
\renewcommand{\thefigure}{S\arabic{figure}}

\setcounter{table}{0}
\renewcommand{\thetable}{S\arabic{table}}

\setcounter{equation}{0}
\renewcommand{\theequation}{S\arabic{equation}}

\section*{Supplementary Information}

\vspace{2em}

\section{Potts reconstruction across \AtwoA{} ensembles}
\label{sec:supp_marginals}

The main text establishes, using the INapo ensemble, that a Potts interaction graph built with a 6~\AA{} $C_\alpha$ cutoff provides a strong compromise between locality and statistical fidelity
(Fig.~\ref{fig:locality_validation}).
Supplementary analyses extend this validation in two directions.
First, the same cutoff dependence is reproduced for FApoG, demonstrating that the 6~\AA{} choice is not specific to the inactive apo ensemble used in the main text.
Second, the specific 6~\AA{} Hamiltonians used in the later \PHASE{} analyses reproduce the MD statistics of the ensembles on which they are fitted.

\begin{figure}[ht]
    \captionsetup{font={small,stretch=1.0}}
    \centering
    \begin{subfigure}[t]{\textwidth}
        \centering
        \includegraphics[width=\linewidth]{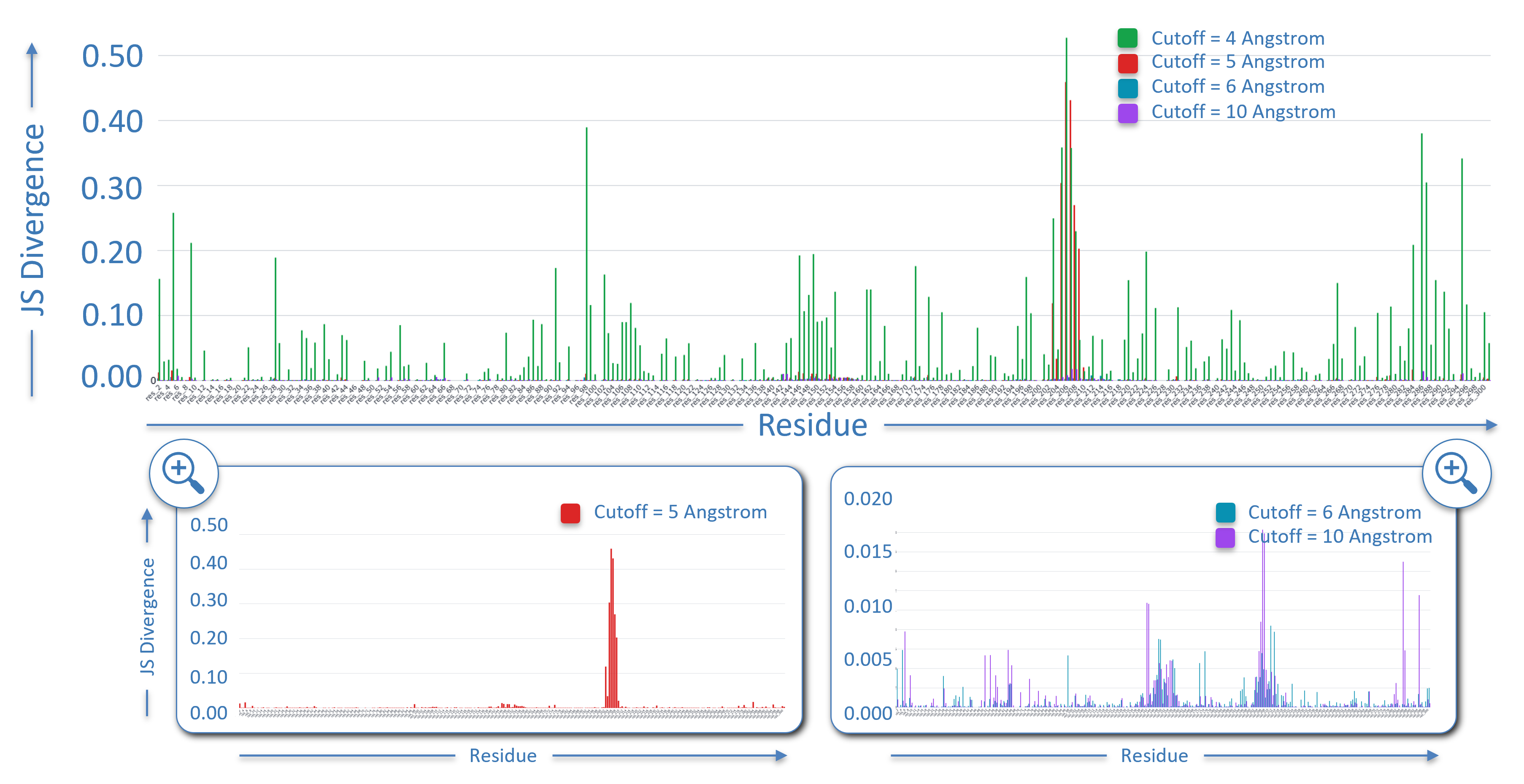}
        \caption{\textbf{Residue-wise marginal distributions.}
        Jensen--Shannon divergence between \textbf{FApoG} residue microstate
        populations measured in MD and in Potts samples for
        $C_\alpha$ contact cutoffs of 4, 5, 6 and 10~\AA{}.}
        \label{fig:supp_fapog_nodes}
    \end{subfigure}
    \vspace{0.8em}
    \begin{subfigure}[t]{\textwidth}
        \centering
        \includegraphics[width=0.8\linewidth]{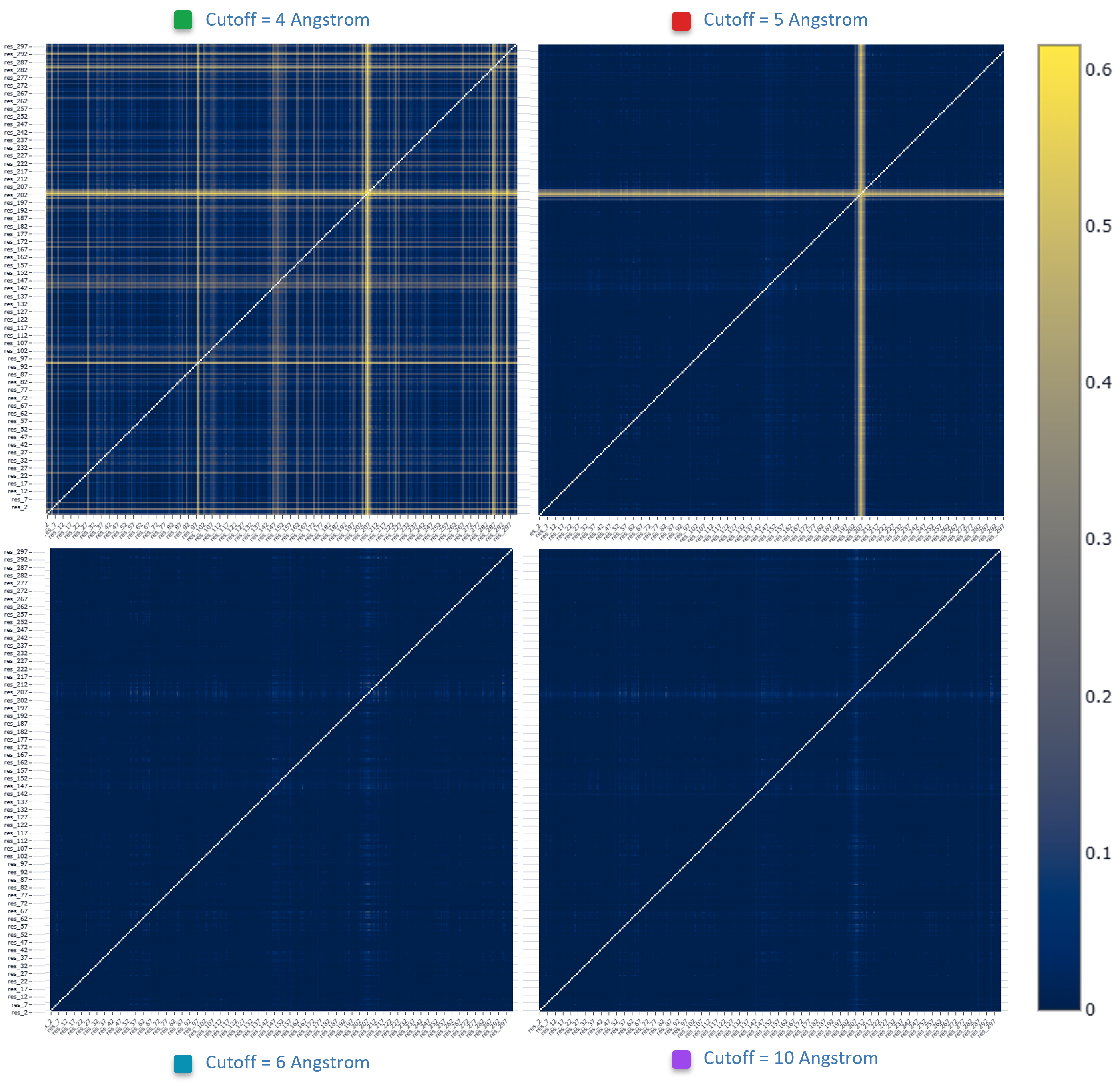}
        \caption{\textbf{All-residue-pair joint distributions.}
        Jensen--Shannon divergence between MD and Potts-sampled joint
        microstate distributions for every residue pair in FApoG.}
        \label{fig:supp_fapog_pairs}
    \end{subfigure}

    \caption{\textbf{Cutoff robustness of local Potts reconstruction in the
    FApoG ensemble.}
    The same locality analysis shown in the main text for INapo is repeated for the active, mini-Gs-coupled FApoG ensemble.
    The 6~\AA{} cutoff selected from INapo generalizes to a distinct active-side ensemble.}

    \label{fig:supp_fapog_cutoff_sweep}
\end{figure}

Supplementary Fig.~\ref{fig:supp_fapog_cutoff_sweep} reports the full cutoff analysis for FApoG, the active-starting apo receptor coupled to mini-Gs.
Potts Hamiltonians are fitted using contact graphs constructed with $C_\alpha$ cutoffs of 4, 5, 6 and 10~\AA{}, and the resulting samples are compared with the MD ensemble through residue-wise and residue-pair Jensen--Shannon (JS) divergences.
As in the main-text INapo analysis, pairwise validation is performed over \emph{all} residue pairs rather than only over pairs connected by explicit Potts couplings.
The same qualitative trend is recovered.
The 4~\AA{} graph is too sparse to reproduce the reduced ensemble accurately, the 5~\AA{} graph improves substantially, and the 6~\AA{} graph already reaches a level of agreement close to that obtained with the denser 10~\AA{} graph.
This supports the use of a common 6~\AA{} cutoff across distinct conformational states of \AtwoA, while leaving open whether the optimal locality threshold is system-specific or transferable to other proteins.

\begin{figure}[ht]
    \captionsetup{font={small,stretch=1.0}}
    \centering
    \includegraphics[width=\textwidth]{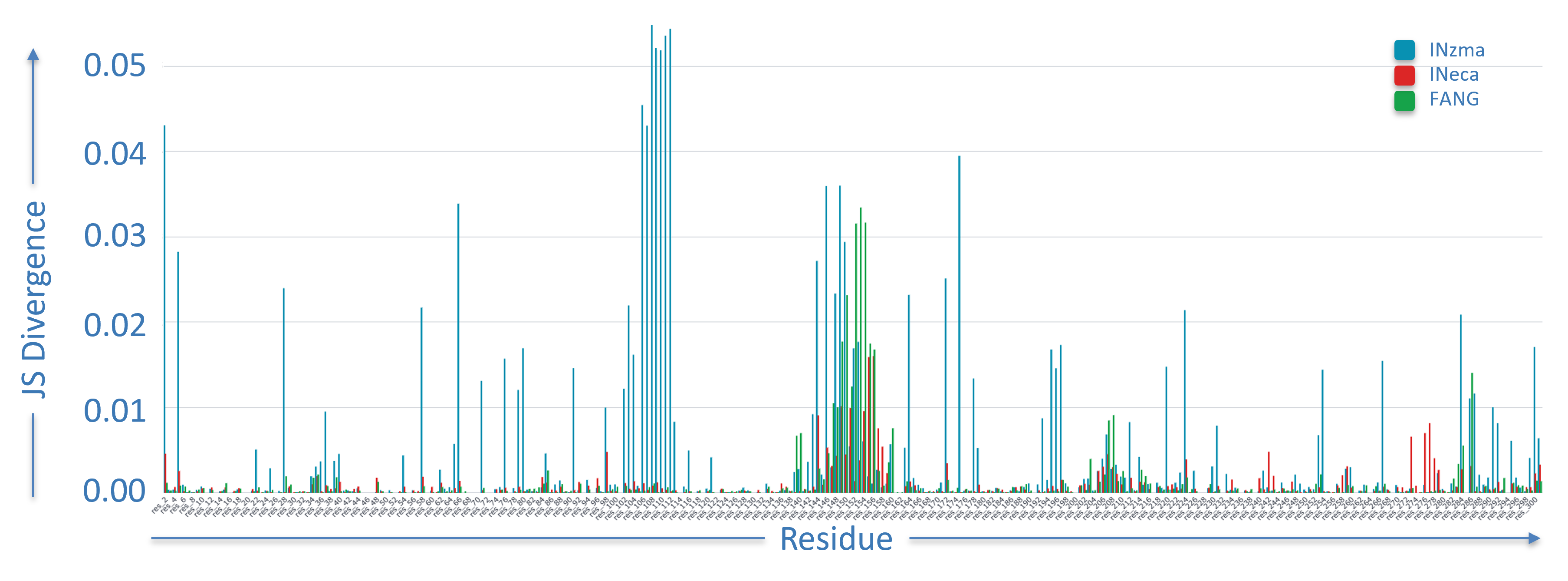}
    \caption{\textbf{Representative validation of the 6~\AA{} cutoff across additional \AtwoA{} ensembles.}
    Residue-wise Jensen--Shannon divergences between MD and Potts-sampled ensembles are shown for the INzma, FANG and INeca Hamiltonians fitted with the selected 6~\AA{} contact graph. The low divergences support the use of the INzma and FANG Hamiltonians for endpoint preference scoring and of the INeca Hamiltonian for Potts sampling followed by all-atom reconstruction.}

    \label{fig:supp_key_models_js_r6}
\end{figure}

Supplementary Fig.~\ref{fig:supp_key_models_js_r6} reports the statistical fidelity of the particular 6~\AA{} Hamiltonians used in the later analyses of this work.
These include the endpoint models INzma and FANG, which define the $\Delta E_{\mathrm{INzma-FANG}}$ preference score, and the INeca Hamiltonian, which is used for Potts sampling in the endpoint ordering analysis in \ref{sec:endpoint_ordering}.
For each system, MD and Potts-sampled ensembles are compared through residue-wise JS divergences.
In all three cases, the selected local Hamiltonian reproduces the residue-level state populations with low divergence.

Together, these supplementary analyses demonstrate that the statistical reconstruction achieved with a 6~\AA{} graph is robust across distinct \AtwoA{} ensembles and that the specific Hamiltonians used for endpoint ordering and generative sampling faithfully reproduce the reduced MD ensembles on which those downstream analyses depend.

\section{Temporal-block generalization in complete reduced configuration space}
\label{sec:supp_temporal_validation}

The marginal analyses above evaluate whether Potts-generated configurations reproduce residue-wise and pairwise statistics of the MD ensembles.
We additionally test whether complete sampled configurations occupy the same region of the reduced configuration space as temporally unseen MD frames.

The 50,000 reduced configurations extracted from the INapo trajectory are
divided into two contiguous temporal blocks,

\begin{equation}
    \mathcal{Q}_1
    =
    \left\{
    \mathbf{x}^{(1)},\ldots,\mathbf{x}^{(25,000)}
    \right\},
    \qquad
    \mathcal{Q}_2
    =
    \left\{
    \mathbf{x}^{(25,001)},\ldots,\mathbf{x}^{(50,000)}
    \right\}.
\end{equation}

A Potts Hamiltonian $H_1$ is fitted using $\mathcal{Q}_1$, with the shared microstate basis and the 6~\AA{} contact graph used throughout the main analysis.
Replica-exchange Gibbs sampling of $H_1$ produces a set of generated configurations denoted by $\mathcal{S}_1$.
The second temporal block $\mathcal{Q}_2$ is not used during fitting and provides a temporally disjoint set of INapo configurations for evaluation.

To compare complete reduced configurations, we define a distance directly from the zero-sum-gauge parameters of $H_1$.
For two configurations $\mathbf{x}$ and $\mathbf{y}$, the single-residue contribution is
\begin{equation}
D_h(\mathbf{x},\mathbf{y})
=
\frac{1}{Z_h}
\sum_{r=1}^{N}
\left|
\widetilde h_r(x_r)
-
\widetilde h_r(y_r)
\right|,
\label{eq:supp_nn_field_distance}
\end{equation}
where
\begin{equation}
Z_h
=
\sum_{r=1}^{N}
\left[
\max_k \widetilde h_r(k)
-
\min_k \widetilde h_r(k)
\right].
\label{eq:supp_nn_field_norm}
\end{equation}
The corresponding pairwise contribution is
\begin{equation}
D_J(\mathbf{x},\mathbf{y})
=
\frac{1}{Z_J}
\sum_{(r,s)\in\mathcal{E}}
\left|
\widetilde J_{rs}(x_r,x_s)
-
\widetilde J_{rs}(y_r,y_s)
\right|,
\label{eq:supp_nn_coupling_distance}
\end{equation}
with
\begin{equation}
Z_J
=
\sum_{(r,s)\in\mathcal{E}}
\left[
\max_{k,\ell}\widetilde J_{rs}(k,\ell)
-
\min_{k,\ell}\widetilde J_{rs}(k,\ell)
\right].
\label{eq:supp_nn_coupling_norm}
\end{equation}
The complete distance is the equally weighted average
\begin{equation}
D(\mathbf{x},\mathbf{y})
=
\frac{
D_h(\mathbf{x},\mathbf{y})
+
D_J(\mathbf{x},\mathbf{y})
}{2}.
\label{eq:supp_nn_total_distance}
\end{equation}

This metric differs from a simple Hamming distance because differences between microstates are weighted by their corresponding field and coupling values.
Consequently, two different residue states contribute weakly when they have similar fitted statistical weights, whereas differences between states associated with more distinct parameters contribute more strongly.
Pair-state differences are treated analogously.
The normalizations in Eqs.~\eqref{eq:supp_nn_field_norm} and \eqref{eq:supp_nn_coupling_norm} place the field and coupling contributions on comparable scales.

For each query configuration $\mathbf{x}$, its distance from the reference ensemble $\mathcal{Q}_1$ is defined as the distance to the closest configuration contained in that ensemble,
\begin{equation}
D_{\min}(\mathbf{x};\mathcal{Q}_1)
=
\min_{\mathbf{y}\in\mathcal{Q}_1}
D(\mathbf{x},\mathbf{y}).
\label{eq:supp_nn_min_distance}
\end{equation}
The analysis therefore does not require a frame-by-frame correspondence between generated and MD configurations.
Instead, each query configuration is compared with the closest configuration observed in the reference temporal block.

Using the same distance definition and reference ensemble $\mathcal{Q}_1$, we evaluate the distributions of $D_{\min}(\mathbf{x};\mathcal{Q}_1)$ for Gibbs samples $\mathcal{S}_1$ generated from $H_1$, the temporally held-out INapo block $\mathcal{Q}_2$, Gibbs samples generated from the Hamiltonian fitted to the complete INapo trajectory, and the pAs and FApo MD ensembles (Supplementary Fig.~\ref{fig:supp_temporal_validation}).

The Gibbs samples generated from $H_1$ occupy a distance distribution comparable to that of the temporally disjoint INapo block $\mathcal{Q}_2$.
The Gibbs samples obtained from the Hamiltonian fitted to the complete INapo trajectory provide an additional comparison with the model used in the main sampling analyses.
In contrast, the pAs and FApo ensembles occupy progressively more distant regions of the reduced configuration space relative to the inactive INapo reference.
Together, these comparisons provide a configuration-level validation complementary to the marginal reconstruction analyses reported in the main text.

\begin{figure}[ht]
    \captionsetup{font={small,stretch=1.0}}
    \centering
    \includegraphics[width=\textwidth]{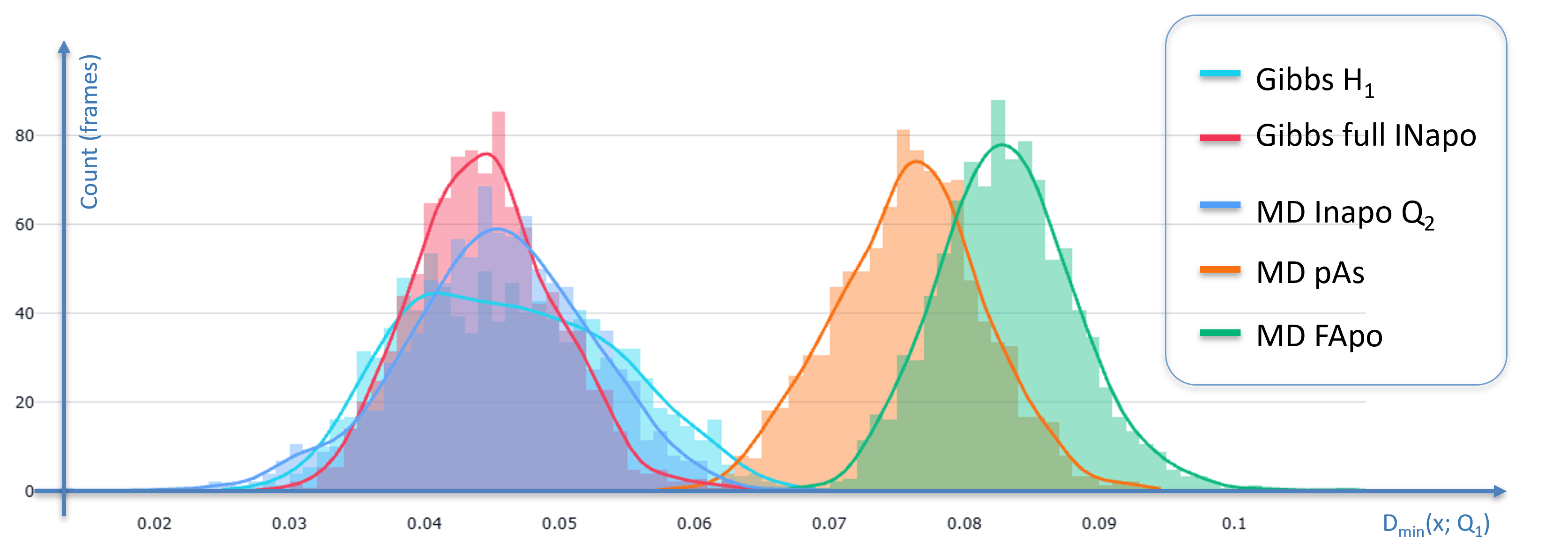}
    \caption{\textbf{Temporal-block validation of complete reduced configurations.}
    Distributions of the nearest-reference distance $D_{\min}(\mathbf{x};\mathcal{Q}_1)$ for Gibbs samples generated from the Hamiltonian $H_1$ fitted to the first temporal block of the INapo trajectory, the temporally held-out INapo block $\mathcal{Q}_2$, Gibbs samples generated from the Hamiltonian fitted to the complete INapo trajectory, and the pAs and FApo MD ensembles. All query configurations are compared with the same reference ensemble $\mathcal{Q}_1$, and distances are computed using the Potts parameters of $H_1$.}
\label{fig:supp_temporal_validation}
\label{fig:temporal_block_validation}

    \label{fig:supp_temporal_validation}
\end{figure}



\section{Validation of Gibbs--Replica Exchange sampling against MD ensembles}
\label{subsec:gibbs_md_validation}

The energetic analyses reported in the main text (Fig.~\ref{fig:regional_endpoint}) are computed from configurations generated by Gibbs--Replica Exchange sampling of the learned Potts Hamiltonians.
To verify that these generated ensembles preserve the statistical organization observed in the original atomistic simulations, we compare endpoint-score distributions obtained from sampled configurations with those computed directly from MD frames.

For each ensemble, we perform Gibbs--Replica Exchange sampling using the corresponding fitted Hamiltonian, generating new discrete residue-state configurations independently from the MD trajectories.
These configurations are evaluated using the inactive and active endpoint Hamiltonians exactly as described for the MD analysis.
The resulting score distributions are then compared with the distributions obtained by evaluating the original MD frames.

\begin{figure}[ht]
    \captionsetup{font={small,stretch=1.0}}
    \centering
    \includegraphics[width=\textwidth]{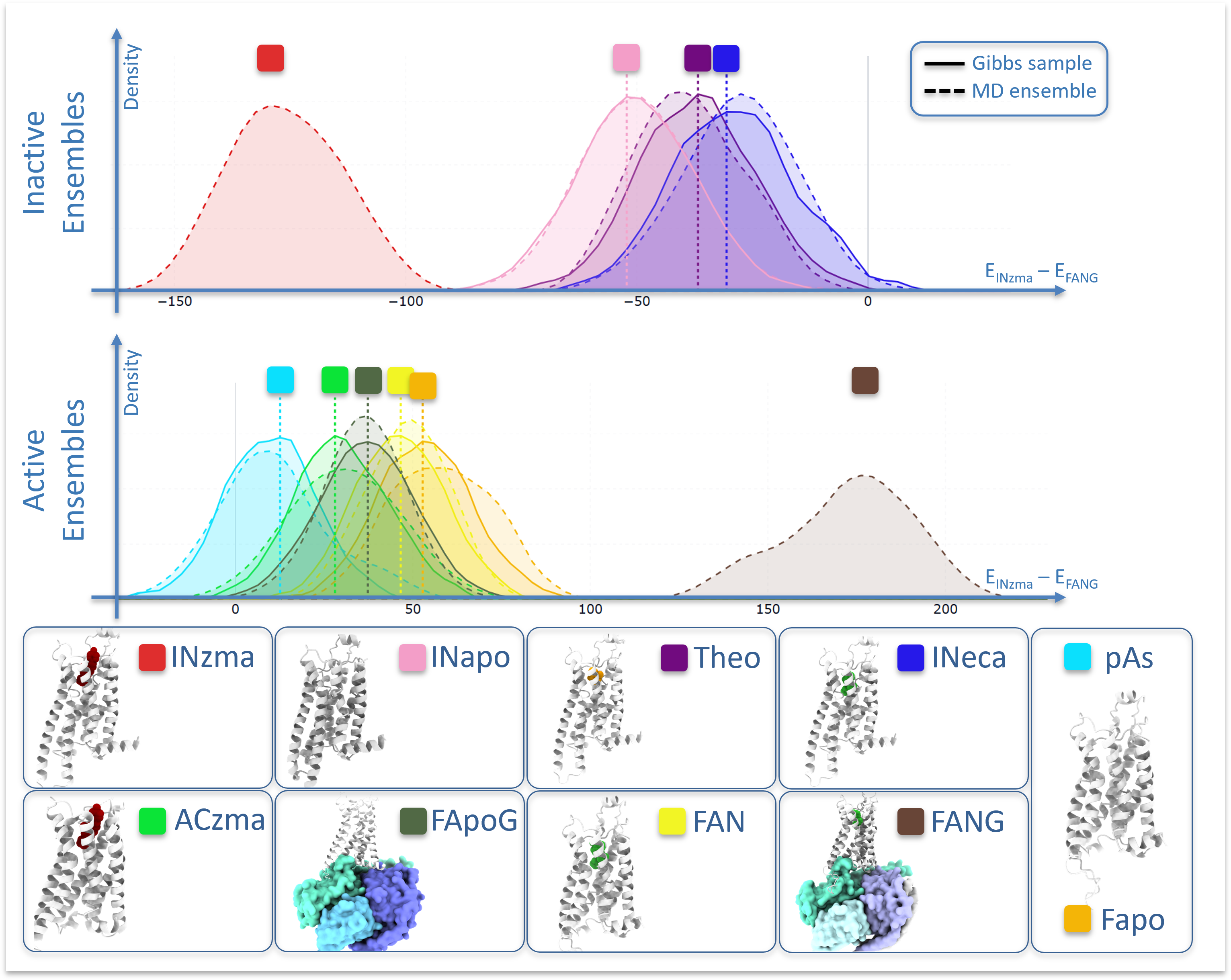}
    \caption{\textbf{Replica-exchange sampling preserves the MD endpoint-score ordering across \AtwoA{} ensembles.}
    Distributions of the endpoint preference score $\Delta E_{\mathrm{INzma-FANG}}$ are shown for replica-exchange Gibbs samples from the fitted 6~\AA{} \PHASE{} Hamiltonians (solid lines) and for the corresponding reduced MD ensembles (dashed lines).
    Inactive-side ensembles are arranged in the top row and active-side ensembles in the bottom row.
    For each ensemble, the sampled distribution closely follows the corresponding MD distribution, preserving the relative ordering across the inactive--active landscape.
    This comparison provides the reference ground truth for the sampled energetic analyses reported in
    Fig.~\ref{fig:regional_endpoint}.}
    \label{fig:supp_md_vs_sampling_energies}
\end{figure}

Figure~\ref{fig:supp_md_vs_sampling_energies} shows this comparison for all analyzed ensembles.
Sampled distributions closely reproduce the corresponding MD energetic profiles across the inactive--active preference coordinate, including ensemble ordering, relative separations and distribution widths.
Agreement is observed both for inactive-like ensembles (INzma, INapo, Theo, INeca) and for active-like ensembles (FApo, pAs, FApoG, FAN and FANG).

These results provide a complementary validation of the generative capability of the learned Hamiltonians.
The Gibbs sampler generates new configurations that preserve the global statistical organization captured by the endpoint-score coordinate.
The energetic analyses presented in the main text can therefore be interpreted as properties of the learned equilibrium distributions and not simply a direct re-evaluations of MD frames.

For the analysis shown in Supplementary
Fig.~\ref{fig:supp_temporal_validation}, all query configurations are
compared with the same reference ensemble, corresponding to the first
temporal block $\mathcal{Q}_1$ of the INapo trajectory. The Potts-weighted
distance is defined using the parameters of $H_1$, fitted to
$\mathcal{Q}_1$. We evaluate $D_{\min}(\mathbf{x};\mathcal{Q}_1)$ for Gibbs
samples generated from $H_1$, the temporally held-out INapo block
$\mathcal{Q}_2$, Gibbs samples generated from the Hamiltonian fitted to the
complete INapo trajectory, and the pAs and FApo MD ensembles.

\end{document}